\documentclass[sigconf]{acmart}
\usepackage{graphicx}
\usepackage{subfig}
\usepackage{textcomp}
\usepackage{multirow}
\usepackage{framed}
\usepackage{amsmath}
\usepackage{listings}
\usepackage{comment}
\usepackage{microtype}
\usepackage[noend]{algpseudocode}
\usepackage[nothing]{algorithm}
\PassOptionsToPackage{hyphens}{url}
\usepackage{hyperref}
\usepackage{enumitem}
\usepackage{tabularx}
\usepackage[normalem]{ulem}

\usepackage{amssymb}
\usepackage{pifont}

\gappto{\UrlBreaks}{\UrlOrds}
\hypersetup{breaklinks=true}
\input{macros}
\renewcommand\footnotetextcopyrightpermission[1]{}
\setcopyright{none}

\begin{document}

\title{MQFQ-Sticky: Fair Queueing For Serverless GPU Functions}

\date{}

\author{Alexander Fuerst}
\affiliation{
  \institution{Indiana University}
  \city{Bloomington,IN}
  \country{USA}
}
\email{alfuerst@iu.edu}

\author{Siddharth Anil}
\affiliation{
  \institution{Indiana University}
  \city{Bloomington, IN}
  \country{USA}
}
\email{sidanil@iu.edu}

\author{Vishakha Dixit}
\affiliation{
  \institution{Indiana University}
  \city{Bloomington, IN}
  \country{USA}
}
\email{vidiksh@iu.edu}

\author{Purushottam (Puru) Kulkarni}
\affiliation{
  \institution{Indian Institute of Technology, Bombay}
  \city{Mumbai}
  \country{India}
}
\email{puru@cse.iitb.ac.in}

\author{Prateek Sharma}
\affiliation{
  \institution{Indiana University}
  \city{Bloomington, IN}
  \country{USA}
}
\email{prateeks@iu.edu}

\begin{abstract}
Hardware accelerators like GPUs are now ubiquitous in data centers, but are not fully supported by common cloud abstractions such as Functions as a Service (FaaS).
Many popular and emerging FaaS applications such as machine learning and scientific computing can benefit from GPU acceleration.
However, FaaS frameworks (such as OpenWhisk) are not capable of providing this acceleration because of the impedance mismatch between GPUs and the FaaS programming model, which requires virtualization and sandboxing of each function.
The challenges are amplified due to the highly dynamic and heterogeneous FaaS workloads. 

This paper presents the design and implementation of a FaaS system for providing GPU acceleration in a black-box manner (without modifying function code).
Running small functions in containerized sandboxes is challenging due to limited GPU concurrency and high cold-start overheads, resulting in heavy queueing of function invocations. 
We show how principles from I/O scheduling, such as fair queuing and anticipatory scheduling, can be translated to function scheduling on GPUs. 
We develop MQFQ-Sticky, an integrated fair queueing and GPU memory management approach, which balances the tradeoffs between locality, fairness, and latency.
Empirical evaluation on a range of workloads shows that it reduces function latency by $2\times$--$20\times$ compared to existing GPU and CPU queueing policies.



\end{abstract}

\maketitle
\pagestyle{plain}

\section{Introduction}

Function as a Service (FaaS) has emerged as a common, narrow interface for a wide range of cloud-native applications, and has become an important abstraction and workload for cloud providers. 
Users are enticed by its dynamic scaling, low cost, and ease of management, since the lifecycle of self-contained \emph{functions} is fully orchestrated by the FaaS provider. 
For FaaS providers (such as Amazon Lambda, Globus Compute~\cite{funcx_hpdc_20}, and others), the major challenge is to efficiently and safely run heterogeneous serverless functions for diverse applications such as machine learning, web serving, IoT, scientific computing, multimedia processing, etc.

FaaS providers must support these diverse workloads, while simultaneously providing low latency and increasing their cluster hardware utilization.
This has been addressed by using standard operating system scheduling and containerization mechanisms for resource control and performance isolation, and a spectrum of resource management policies for scheduling, load-balancing, queueing, and keep-alive. 
This has allowed FaaS providers to run hundreds of functions concurrently on a single server, each with extremely high heterogeneity in terms of their execution time, arrival rate, memory consumption, etc~\cite{shahrad2020serverless, bauer2024globus}.

Conventionally, serverless function execution has been restricted to CPUs.
However, GPU accelerators are now a common feature in data center and edge servers. 
Functions are increasingly amenable to GPU acceleration, due to the adoption of the FaaS abstraction by computationally intensive workloads such as machine learning~\cite{carreira2018case,romero2021llama,gimeno2022mlless,xu2021lambdadnn}, scientific computing~\cite{kumanov2018serverless,hung2019rapid, aytekin2019harnessing,werner2018serverless,shankar2020serverless}, multimedia~\cite{ao2018sprocket, zhang2019video}, etc.
Among public clouds, to the best of our knowledge, only Alibaba provides serverless GPU functions~\cite{alibaba-gpu-function}.
Opensource FaaS frameworks such as OpenWhisk, which are used for many cloud systems such as Funcx~\cite{funcx_hpdc_20} do not support GPU acceleration.

\emph{Can we retain the conveniences of the FaaS abstraction and still provide efficient GPU execution for highly heterogeneous and dynamic FaaS workloads?}
This is challenging because GPU resource management for serverless functions is a fundamentally different problem compared to existing techniques for ``CPU functions''. 

Due to their hardware architecture and programming model, GPU multiplexing is significantly inferior to CPUs.
GPUs are designed for ``batch processing'' of large computational kernels, whereas CPUs can run hundreds of functions concurrently with processor-sharing. 
Black-box resource multiplexing (i.e., without modifying the function) is the key building-block of serverless computing, and our first challenge and contribution is to improve the latency of \emph{containerized} functions with GPU multiplexing.
Since GPU memory is limited, we develop new memory management techniques integrated with function execution. 
Our memory prefetching and swapping are implemented using CUDA's Unified Virtual Memory (UVM~\cite{nvidia-uvm}), and provide the foundation for practical use of GPUs for serverless functions.

In spite of GPU-level optimizations, the limited multiplexing and run-to-completion execution model results in significant \emph{queueing}, as functions wait for GPU resources. 
\emph{The main focus and primary contribution of this paper is in developing GPU scheduling policies.}
We observe that cold-starts and (temporal) locality play a vital role in GPU-function performance. 
Cold-starts associated with initializing GPU containers (such as NVIDIA-Docker~\cite{nvidia-container}) can take \emph{several seconds} and increase latency by up to $100\times$, and have also been observed in Alibaba's GPU functions~\cite{alibaba-gpu-noshare}. 
While batching is a common technique with GPUs~\cite{ali2022optimizing}, doing so in a black-box and \emph{fair} manner is challenging.
This is compounded by FaaS workloads where the execution time, memory requirements, and arrival rates can all vary by orders of magnitude~\cite{shahrad2020serverless}.

As a result of these key differences, current FaaS scheduling developed for CPUs (e.g., OpenWhisk uses FCFS for queueing functions if CPU resources are unavailable~\cite{kaffes2021practical}), can lead to excessive queueing, and can increase the end-to-end latency several-fold, as we empirically demonstrate.
To address these challenges, we build on the foundations provided by \emph{fair queueing}~\cite{goyal1997start}, to provide equal/fair access to GPU resources for heterogeneous functions. 
However, standard fair queueing is not locality-optimized, as also recently observed in the field of LLM inference~\cite{cao2025locality}. 

Our second main insight is that the GPU function scheduling problem is close to \emph{I/O scheduling}, where locality plays a similarly important role. 
Inspired by MQFQ (Multi-Queue Fair Queuing~\cite{hedayati2019multi}) and anticipatory scheduling~\cite{iyer2001anticipatory}, we develop the \emph{MQFQ-Sticky} GPU function scheduling policy.
MQFQ-Sticky treats each function as a separate application, and dispatches invocations in order of their service requirements---retaining the fairness properties of MQFQ while significantly reducing cold starts. 
It also provides simpler and more ``interpretable'' control knobs for users and cluster operators. 
Our scheduling policies provide a unified interface to using GPU multiplexing capabilities such as MPS~\cite{nvidia-mps} and MIG~\cite{nvidia-mig, li2022miso}, and also allows FaaS operators to harness older GPUs without these hardware features.

Developing an equivalence between I/O and GPU scheduling allows us to leverage classic and well studied ideas from disk scheduling, and provides a new and more principled approach to GPU resource management. 
Our work is unique in that it seeks to improve GPU utilization through scheduling, and does not depend on specialized application level approaches or hardware virtualization support. 
A major focus of prior work on GPU support for FaaS has been on application-specific optimizations (such as for ML inference like Paella~\cite{ng2023paella} and others~\cite{gujarati2020serving}) or disaggregation mechanisms, which are not directly applicable to black-box functions.

FaaS occupies the challenging middle-ground between ML-inference and long-running batch applications, and combined with its high sensitivity to locality, presents new challenges to efficient GPU use. 
As GPUs become ubiquitous and less ``special'', providing best-effort acceleration in a scalable and flexible manner to diverse applications becomes increasingly important for cloud platforms. 
To the best of our knowledge, this is the first work to provide local GPU acceleration for FaaS, and we make the following contributions: 
\begin{enumerate}
\item We develop fair-queuing based scheduling (called \quotes{\QName}) for black-box containerized functions, which preserves both locality and fairness by adapting ideas from I/O scheduling.
This allows us fine-grained and intuitive control of both temporal and spatial multiplexing of GPU compute and memory resources.
Our GPU memory management is integrated with the scheduler, and by virtually eliminating cold-starts, provides more than $300\times$ reduction in latency compared to current unoptimized GPU containers.

\item We empirically evaluate our scheduling policies and their tradeoffs using a range of FaaS workloads. Compared to other scheduling policies such as continuous batching or Paella's~\cite{ng2023paella} \quotes{fair SJF}, \QName~reduces the average latency by $1.2\times$--$20\times$ and the variance by $3\times$--$8\times$. 
  
\item  MQFQ-Sticky is a general-purpose GPU scheduling approach and works with MPS, MIG, and scales to multiple GPUs. We provide the first open-source, practical, and high-performance GPU acceleration for black-box functions. 
  
\end{enumerate}

\vspace*{-0.2cm}



y
\section{Background and Motivation}
\label{sec:bg}


Serverless computing entails executing snippets of user code, usually in an event-driven manner, inside protected and isolated sandboxes~\cite{serverless-cacm-21}.
For users, serverless functions offer many features such as elastic scaling and scale-to-zero, which makes cloud-native applications easier to develop and deploy. 
The FaaS \emph{provider} usually runs a \emph{control plane} such as OpenWhisk and OpenFaaS (or proprietary ones in the case of popular services like Amazon Lambda~\cite{aws-lambda}), for handling the scheduling, scaling, load-balancing, resource limiting, and accounting for each invocation. 

\subsection{Function Execution and Scheduling}

In general, functions are run inside containerized sand-boxes (such as Docker~\cite{docker-main} or lightweight VMs~\cite{firecracker-nsdi20}).
The placement, scheduling, and execution of containers is orchestrated by the FaaS control plane. 
Because of the richness of their usecases, function workloads are highly diverse with a range of several orders of magnitude in all dimensions.
For example, both Azure~\cite{shahrad2020serverless} and Alibaba~\cite{luo2021characterizing} workloads indicate that the inter-arrival-times can range from 0.1s to hours, the execution times can range from milliseconds to minutes, and the memory footprint from 100 MB to 5 GB. 
From the FaaS control plane's perspective, the popularity and diversity of FaaS workloads makes resource management tasks particularly challenging, and can add several dozen milliseconds of latency in orchestrating function invocations~\cite{cvetkovic2024dirigent, fuerst2023iluvatar, serverless-cluster-cost}.

Due to this workload diversity, scheduling of functions even on a single server is challenging and an active area of research, and a broad spectrum of techniques have been developed for CPU functions.
Based on server level resource availability and function SLA (service level agreement), vertical and horizontal auto-scaling approaches can be effective~\cite{Anubis_2024}. 
The overcommittment of CPU and memory resources is also controllable through queueing: invocations can be enqueued before running inside their containers.
Minimizing function cold-starts is a major pervasive objective across FaaS, and control planes maintain large keep-alive caches, and also employ locality-based scheduling. 
For instance, the OpenWhisk load-balancer and workers maintain distributed FCFS queues and prioritize functions with pre-existing warm containers. 
Other policies such as Shortest Job First and Earliest Virtual Deadline~\cite{fuerst2023iluvatar, zuk_call_2022} can also be effective for heterogeneous FaaS workloads. 
However, all the scheduling and keep-alive policies have been developed for large number of CPU cores and memory---assumptions incompatible with GPUs.



\subsection{GPU Acceleration for Functions}

\newcommand{\rod}{\cite{che2009rodinia}}

\begin{table}
  \caption{Latencies (in seconds) for GPU and CPU \textbf{W}arm and \textbf{C}old function invocations.}
  \label{tab:gpu-cpu}
  \begin{tabular}{@{}l@{}|@{}cc@{}|@{}cc@{}}
    \hline
    Function & GPU [\textbf{W}] & CPU [\textbf{W}] & GPU [\textbf{C}] & CPU [\textbf{C}] \\
    \hline
  Imagenet [ML] & 2.253 & 5.477 & 11.286 & 10.103 \\
  Roberta [ML] & 0.268 & 5.162 & 15.481 & 14.372 \\
  Ffmpeg [Video] & 4.483 & 32.997 & 4.612 & 34.260 \\
  FFT [HPC] & 0.897 & 11.584 & 3.322 & 13.073 \\
  Isoneural [HPC] &0.026 & 0.501 & 9.963 & 1.434 \\
  Lud \rod & 2.050 & 70.915 & 2.359 & 110.495 \\
  Needle \rod & 1.979 & 144.639 & 2.177 & 223.306 \\
  Pathfinder \rod & 1.472 & 134.358 & 1.797 & 106.667 \\
  \end{tabular}
\end{table}

Many applications that have adopted FaaS for its on-demand scaling also benefit from GPU acceleration, as shown by Table~\ref{tab:gpu-cpu}, which shows the function execution times on a NVIDIA V100 GPU and a Intel Xeon Platinum 8160 (48 core) CPU. 
CPU functions are allocated one CPU core, matching the allocations in public FaaS platforms, and GPU functions can use the entire accelerator.
ML inference tasks such as Imagenet and Roberta see a $3\times$ and $20\times$ reduction in latency compared to a warm CPU container. 
Video encoding via \texttt{ffmpeg}, which is one of the most popular functions on AWS Lambda~\cite{aws-netflix}, can leverage specialized hardware found in most GPUs for a 7x speedup. 
Scientific computing has also started using FaaS~\cite{john_sweep_2019,mocskos_faaster_2018,werner2018serverless,shankar2020serverless}, and benefits from GPU acceleration for its common primitives such as FFT ($13\times$ faster). 

The serverless abstraction allows decoupling of computation from its location, and prior work has investigated the use of remote disaggregated GPUs for FaaS~\cite{naranjo2020accelerated,fingler2022dgsf}, and providing acceleration as a service~\cite{varghese2015acceleration,du2022serverless}.
These prior efforts have focused on the virtualization of accelerators through new mechanisms and abstractions, and often for specialized workloads such as ML inference.
We focus on a higher level of abstraction, and develop a scheduling architecture for providing \emph{local} GPU acceleration. 
\section{GPU-Function Challenges}
\label{sec:motiv}

In this section, we explain the challenges and tradeoffs in GPU function execution, and how resource management techniques developed for CPU functions are incompatible. 



\subsection{Container-level Black-box Multiplexing}

Since FaaS is the common abstraction for supporting a wide range of applications, we seek a general-purpose solution which supports arbitrary functions on heterogeneous clusters, and preserves existing resource isolation guarantees.
FaaS platforms need to run arbitrary user code in containers, without modifying the code or making assumptions about it: i.e., functions are \textbf{black-boxes} from a provider perspective. 
This puts us in a challenging middle-ground between GPU virtualization for VMs and containers, and the plethora of application-specific GPU acceleration middleware. 

GPU virtualization techniques developed for long-running VMs and containers~\cite{suzuki2014gpuvm, yeh2020kubeshare} are \textbf{throughput-oriented}~\cite{zhu_gass_2024}, and not applicable to latency-sensitive dynamic FaaS workloads requiring more concurrency. 
New GPU hardware features for multiplexing such as NVIDIA's MPS (Multi-Process Service~\cite{nvidia-mps}) and MIG (Multi-Instance-GPUs~\cite{nvidia-mig}) can increase multiplexing for containerized applications---however the fundamental execution model is still run-to-completion instead of processor-sharing like with CPUs~\cite{ausavarungnirun2018mask, jang2019heterogeneous}. 
Thus, even with additional hardware support, GPUs are still an order of magnitude away from CPUs in terms of temporal multiplexing frequency and spatial multiplexing capacity.

The programming model and software stacks of GPUs supports and often \emph{requires} application-level performance optimizations, at the cost of portability and abstraction. 
For instance, there has been a plethora of recent work in optimizing the training and inference of ML models on GPUs~\cite{xiao2018gandiva, mahajan2020themis, weng2022mlaas, ye2024deep}. 
Similar to FaaS, ML inference workloads are also highly dynamic and require fine-grained multiplexing. 
These challenges can be overcome by application-level batching~\cite{ali2022optimizing}, GPU kernel preemption~\cite{han2022microsecond}, combining GPU spatial and temporal hardware multiplexing~\cite{li2022miso}, etc.
ML inference is such a dominant workload that several recent works targeting ``serverless GPUs''~\cite{gu2023fast,wu2024streambox} are in fact \textbf{limited to inference tasks and are not general solutions}. 

The run-to-completion execution model and multiplexing limitations result in significant queueing for GPU functions.
These challenges are not faced by conventional CPU functions. 
Resource management for CPU functions is often more ``downstream'' in the invocation path and can be handled using existing OS support. 
For example, dynamic CPU bandwidth control using cgroups is common in function auto-scaling~\cite{ensure-faas-acsos20}, and CPU-scheduler time-slices can be tuned to reduce tail and average latency~\cite{fu2024alps, fu2022sfs}. 
Thus due to processor-sharing and OS support, queueing-overhead for CPU functions is usually under 50\% even for extreme loads~\cite{zuk_call_2022, zuk_scheduling_2020, fuerst2023iluvatar}---whereas for GPU functions, it can increase latency several-fold, as we show in our empirical evaluation. 

The number of concurrently executing functions should also be kept low to mitigate \textbf{performance interference}, which can be excessive from both compute~\cite{yamagiwa2009performance,phull2012interference} and data movement between host and device~\cite{yu2019automatic, hong2017gpu}. 
This reduction in concurrency increases queue waiting times. 
Finally, since function workloads are highly heterogeneous and dynamic, we must select the \quotes{active} functions carefully so as to balance fairness and throughput.

\subsection{Extreme Cold-start Overheads For GPU Containers}

\begin{figure}
  \includegraphics[width=0.45\textwidth]{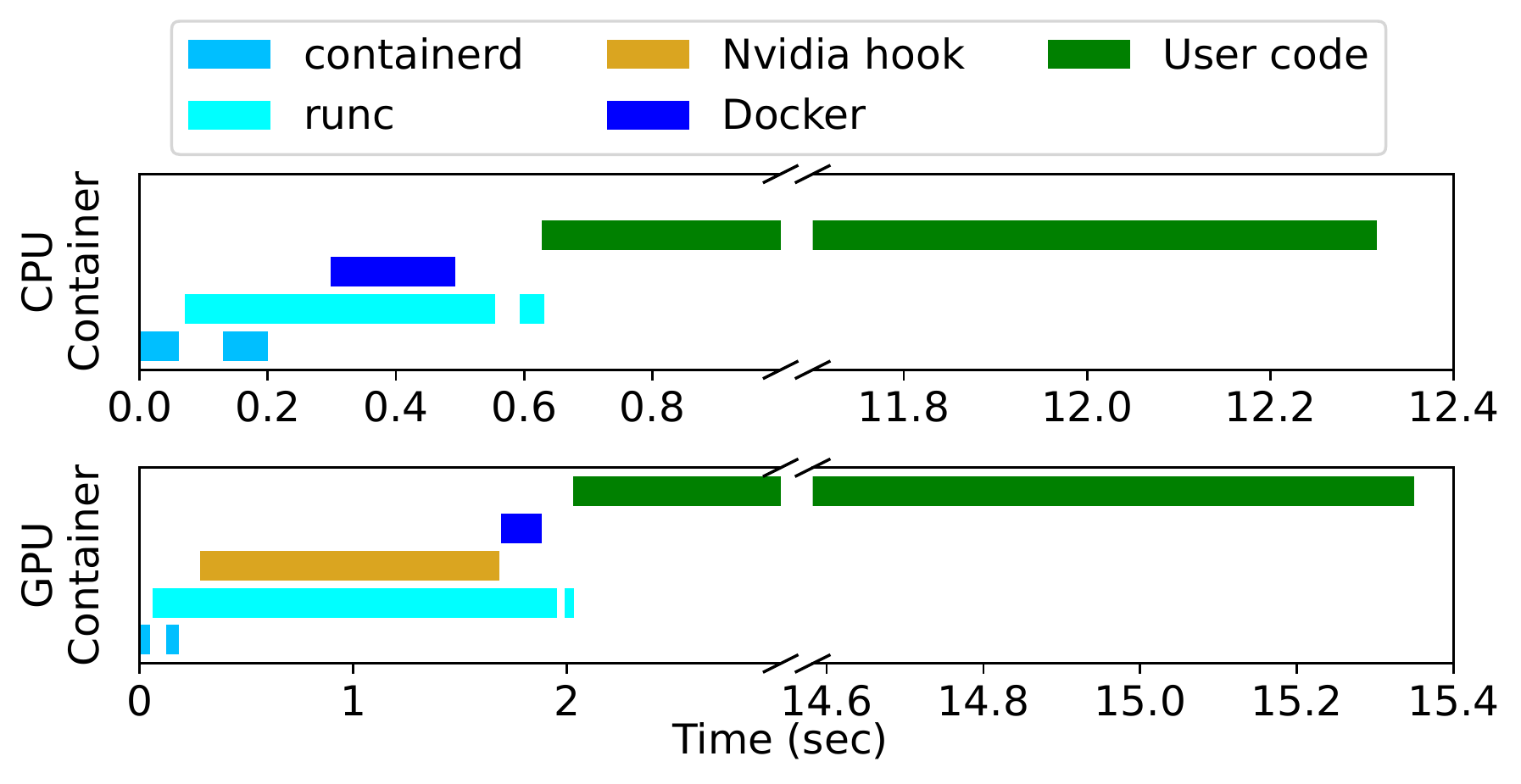}
  \caption{Timeline of cold-starts of CPU (top) and GPU (bottom) function containers running TensorFlow inference code. 
    GPU initialization and code dependencies increase latency by three seconds.}
  \label{fig:cold-timeline}
  \vspace*{-9pt}
\end{figure}

Cold-starts due to sandbox creation and initialization are a well known performance problem for serverless functions~\cite{du2020catalyzer,lin_mitigating_2019,manner_cold_2018,mohan_agile_2019}.
We find that such cold starts are severely exacerbated by GPU containers, increasing latency by up to $75\times$ (Table~\ref{tab:gpu-cpu}).
These delays are so significant that even CPU cold starts can provide lower latency for some functions.
A breakdown and comparison of these overheads for the ML inference function (using TensorFlow) is shown in Figure~\ref{fig:cold-timeline}.
For the GPU container (bottom figure), the Nvidia hook library adds more than 1.5 seconds of delay when it attaches the GPU to the container.
User function code loads additional GPU-specific libraries and dependencies, and its startup requires 1.5 additional seconds.

A common approach to alleviate cold starts is to keep the container warm in memory.
This can be used GPU containers, but introduces additional challenges because of limited GPU memory. 
For example, the NVIDIA V100 has only 16 GB VRAM, compared to terabytes of DRAM available for CPU container keep-alive cache.
Current keep-alive policies for CPU containers use caching-inspired techniques where the working-set is smaller than the available memory~\cite{faascache-asplos21}. 
Thus unlike CPUs, GPU keep-alive has significant opportunity cost, and CPU keep-alive techniques do not directly alleviate the large cold-start overheads.




\section{Design: Scheduling GPU Functions}
\label{sec:design}

The different hardware and workload model of GPUs requires a different approach as compared to conventional CPU function scheduling.
To tackle these challenges, we develop an \textbf{integrated scheduling framework} which combines new solutions for locality-aware fair queueing, proactive memory management, and GPU load management. 
We first show that fair queuing, as used in I/O scheduling, can be a useful framework for GPU scheduling (Section~\ref{sec:d:mqfq}), then describe the GPU-specific scheduling (Section~\ref{sec:mq}) and memory management optimizations (Section~\ref{sec:design-cont-shim}). 

\begin{figure}
  \includegraphics[width=0.4\textwidth]{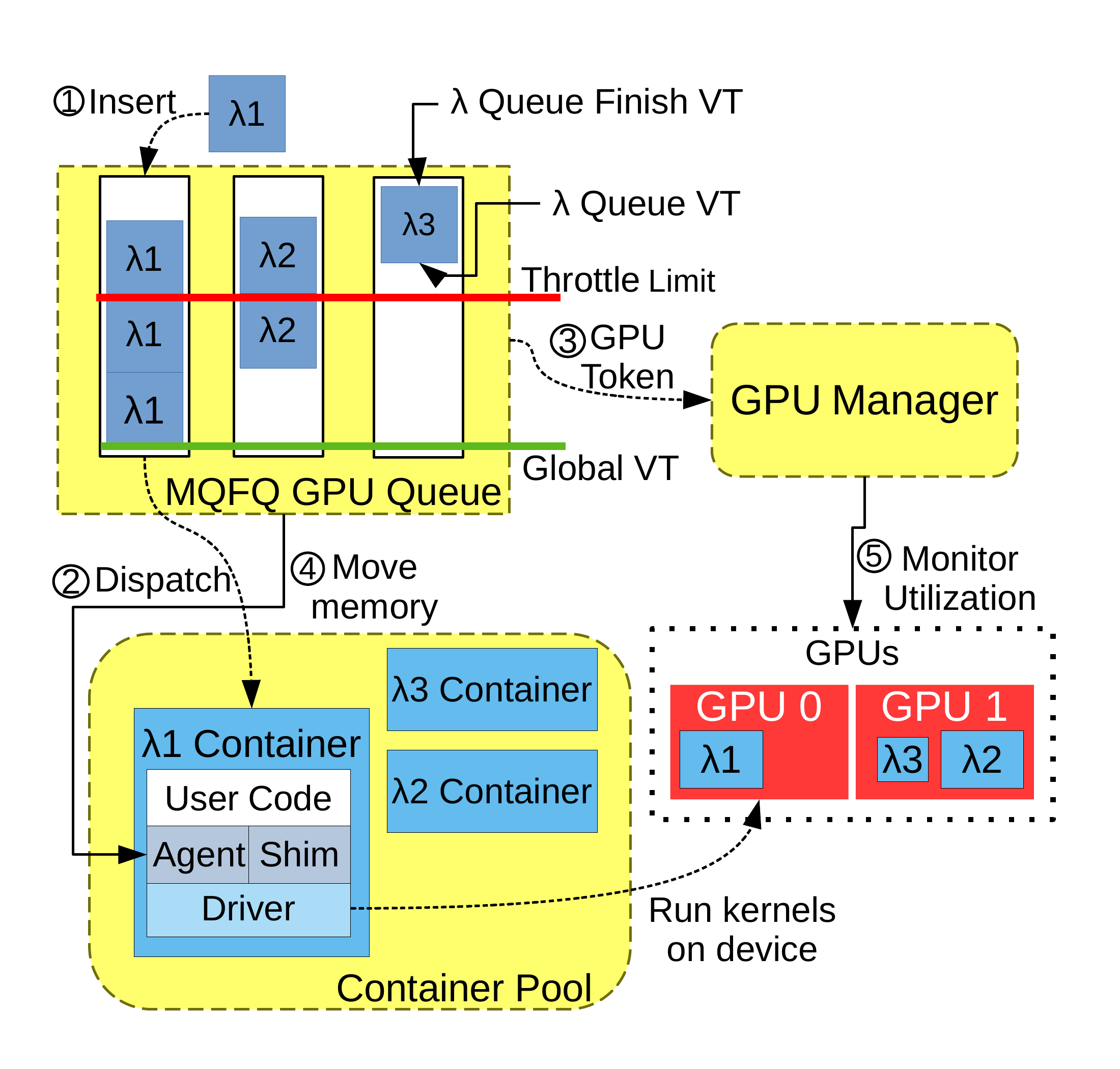}
  \caption{Scheduling GPU functions from individual queues with Multi-Queue Fair Queuing. Invocations are dispatched based on the virtual time. The container pool helps with warm starts.}
  \label{fig:sys-diag}
\end{figure}

\begin{table}[!t]
  \begin{tabular}{c|p{6.8cm}}
    \hline
    Symbol & Description \\
    \hline
    \VT & Virtual Time, the total service time accrued by a queue \\
    \GlobVT & Minimum \VT~across all active queues \\
    \T & Amount any queue's \VT~can exceed \GlobVT~before being throttled \\
    \D & Device concurrent invocations, can be fixed or dynamic with an upper-bound \\
    TTL & Time-to-live for an empty queue to become inactive. Set to $\alpha \times \text{IAT}$ (inter-arrival-time). \\
\hline 
  \end{tabular}
    \caption{Key symbols and parameters for \QName.}
  \label{tab:mq-symbols}
  
\end{table}

\subsection{GPUs as Multi-Queue I/O Devices}
\label{sec:d:mqfq}

We observe that the tradeoffs and challenges of providing fairness and high throughput for serverless GPU workloads are similar to modern I/O scheduling. 
Specifically, we can view GPUs as multi-queue I/O devices, and use fair scheduling algorithms like MQFQ~\cite{hedayati2019multi} to provide a rigorous and well-tested conceptual framework.
\textbf{FaaS and I/O workloads share many similarities}: concurrent applications can also make I/O requests of different sizes (analogous to functions with different GPU service times), and at different rates. 
Most importantly, temporal and spatial locality also plays a central role in I/O. 
Modern disks can handle multiple concurrent dispatches, which also maps to the ability of modern GPUs to be multiplexed across multiple clients.
Because of these similarities, we use I/O scheduling principles as a starting point for GPU-function scheduling.

Our scheduler maintains multiple dispatch queues, each queue corresponding to an individual function (Figure~\ref{fig:sys-diag}). 
Active queues hold pending invocations, and ``inactive'' queues are empty but may be \emph{anticipating} invocations. 
We use the standard fair queuing notion of virtual time (\VT), a monotonically increasing counter which captures the amount of service rendered to queues.
Table~\ref{tab:mq-symbols} shows the key parameters of our fair queueing architecture. 
Queues are selected for dispatch based on their \VT's, and fairness arises from a bound on the maximum difference of queue \VT's. 
A queue's \VT~increases on each dispatch, and each new enqueued invocation gets a virtual start tag based on its queue's \VT~and position in the queue.  
To prevent popular functions from monopolizing the GPU to the detriment of others, queues are \emph{throttled} if their \VT~exceeds the \GlobVT, which is the minimum of all queues' \VT.
In classic fair queueing, the queue with the lowest \VT is picked.
But to increase batching and locality, we allow queues to be dispatched if their \VT is within \T of \GlobVT. 
Invocations from \emph{multiple} active queues can be concurrently dispatched, and this is controlled by the device parallelism parameter \D. 
As we shall show, these parameters provide principled and tunable locality-enhanced fair queueing.
We refer to the original MQFQ paper~\cite{hedayati2019multi} for an expanded algorithmic description of fair queueing.  

\subsection{\QName{}: Locality-enhanced Fair Queuing}
\label{sec:mq}


While there are similarities between I/O and FaaS workloads, there are also \textbf{key differences} which must be considered for scheduling. 
GPU functions are heterogeneous: having different compute and memory footprints, execution runtimes, and cold vs. warm execution times; all of which diverge from disk assumptions. 
The GPU device model is also different: the device parallelism is significantly lower (SSDs support hundreds of active dispatches), and execution performance is highly sensitive to utilization and interference.
Lastly, successive FaaS GPU invocations see lower latency due to temporal and even spatial locality, a fact requiring significant attention when maximizing throughput.
We therefore modify the original MQFQ design to account for these differences, to get the maximum performance out of our accelerators.

\mhead{Architecture}
Our MQFQ-based scheduler can be composed with and layered on top of GPU hardware multiplexing mechanisms such as MPS (Multi-Process Service~\cite{nvidia-mps}) and MIG (Multi-Instance GPU~\cite{nvidia-mig}). 
We maintain a single MQFQ GPU queue per server, and support multiple physical or virtual GPUs (Figure~\ref{fig:sys-diag}), taking advantage of the hardware optimizations for spatial multiplexing when available. 
Our scheduling policy drains the per-function queues and dispatches the invocations to a GPU based on the device load and function locality. 
In the base case without hardware multiplexing support (on older GPUs), we dispatch multiple invocations concurrently.
For GPUs with MPS support, we launch a container hosting the MPS daemon before any functions are run, and connect all future function containers to it, allowing MPS to manage device concurrency. 
In the case of Multi-Instance GPUs (MIG), each instance is treated as a separate virtual GPU, and we dispatch only one function per vGPU. 



\mhead{Per-function Fairness}
Each queue holds requests of different sizes, which is based on the function execution times.
We track the historical average execution time $\tau_k$ of each function $k$, and when an item is dispatched, increment its queue's \VT~by $\tau_k$.
Thus, shorter functions are allowed more \emph{invocations} than their long counterparts, but both get equivalent wall-clock time on the GPU.
After an invocation is dispatched, \GlobVT~is updated if necessary to a potentially new global minima across queue \VT{}s. 

\mhead{Container Warm-pool}
We maintain a small pool of warm GPU containers. 
Since GPU memory is limited, we use the queue states for \textbf{proactive memory management.}
Queues that become active have their data moved onto the device in anticipation of use (if space is available). 
When a container is about to execute, we proactively move all its data to the GPU. 
Conversely, throttled and inactive queues have their containers marked for eviction from the container pool, and are asynchronously moved to CPU memory using an LRU eviction scheme. 
More details about the data monitoring and movement are described in Section~\ref{sec:design-cont-shim}. 

\mhead{Anticipatory Scheduling}
Function performance is impacted by the availability of a warm container and data in GPU memory.  
We introduce anticipatory scheduling to MQFQ to maximize the use of both. 
Anticipatory scheduling for disks~\cite{iyer2001anticipatory} boosts locality by keeping request streams \quotes{active} even if they are empty, in anticipation of future requests, which is especially beneficial for interactive applications.  
If a queue is empty (i.e., it has no pending invocations), then instead of immediately marking it inactive, we provide a grace period.
Without this grace period, because of the proactive memory management described above, functions would see their warm containers immediately removed from GPU memory.
Instead, we keep empty queues active for a configurable TTL (time to live), based on the function's inter-arrival-times.
Specifically, we set the queue TTL to $\alpha \times \text{IAT}$, where $\alpha$ is a tunable parameter. 
This policy is guided by the observation that reuse-distance is long-tailed~\cite{faascache-asplos21}, so a single global TTL is not ideal for both popular and rare functions. 

\mhead{Batching}
Our second technique for improving warm starts is to allow the queues to dispatch invocations in small \quotes{mini-batches}.
An active queue's start time is allowed to be up to \T~units ahead of \GlobVT.
\T~is the second main configurable parameter: larger values will result in larger batches and more locality, but more ``out of order dispatch''.
\T~is thus the 'queue over-run' parameter. 
If $queue.\VT + \T \ge \GlobVT$, then the queue is \emph{throttled}. 
It may return to the \emph{active} state only after other queues get to run and the \GlobVT~increases.

\mhead{Device Concurrency and Load Control}
Because each function uses different amounts of compute and memory during execution, a fixed level of device parallelism (\D) like in disk scheduling may be sub-optimal.
We therefore track memory usage of running containers and GPU utilization to adjust \D~dynamically, to minimize contention and execution overhead.
This utilization-based feedback permits different scheduling rates based on the dynamic workload characteristics.
We take two input parameters: the device utilization threshold (such as 90\%), and the maximum parallelism level (irrespective of utilization).
A thread monitors real-time utilization and changes the \D~level dynamically to ensure the utilization is under the threshold.
Higher thresholds increase utilization and reduce queuing, but risk performance interference. 
More details of memory usage and GPU monitoring are described in Section~\ref{sec:design-cont-shim} and Section~\ref{sec:gpu-man} respectively. 

\mhead{Preferential Queue Dispatch}
\emph{Our insight behind MQFQ-Sticky's fair queueing policy is that the queue over-run parameter $T$ provides additional opportunities for out-of-order execution and improves locality and end-to-end latency.} 
In classic fair queueing and even MQFQ, the queue with the lowest \VT~is always chosen.
However, we can select the next queue for dispatch based on locality, as long as $queue.\VT < \GlobVT + \T$.

Because FaaS workloads can be very heavy-tailed, there is high likelihood of unpopular and rare functions to face excessive queueing delays.
Note that fair queueing only guarantees that each function gets an equal share of \emph{GPU service time}, and not the total end-to-end latency.
Thus, MQFQ-Sticky's heuristic (described next and in Algorithm~\ref{algo:dispatch}) is designed to improve locality as well as reduce the tail latency of functions.

We first filter based on the over-run threshold $T$ to get the candidate queues  (line~\ref{lst:line:filter}). 
Next, we prioritize functions with longer queues, which provides more batching opportunities and reduces their larger backlog. 
Ties are broken in favor of the queue with the least number of currently executing invocations (Line~\ref{lst:line:in_flight}). 
This encourages multiple queues to progress and reduces the chance of a cold start caused by concurrent execution of the same function.
Queue stickiness from this heuristic provides sufficient temporal locality between active queues to maximize throughput.
This completes the description of the key attributes of our MQFQ-Sticky algorithm.

\mhead{Fairness Guarantees}
Through our careful adaptation of MQFQ, we are able to retain its fairness properties, which provide an upper-bound on the difference in GPU service times across any two functions. 
Specifically, let $S$ be the total GPU execution time of backlogged function during the time-span $(t_1, t_2)$.
Recall that a queue is backlogged if it is non-empty.
Based on the main theorem in~\cite{hedayati2019multi}, for all backlogged functions $i$ and $j$, we get:
\begin{equation}
  \label{eq:fairness}
 \left|\dfrac{S_i}{w_i} - \dfrac{S_j}{w_j} \right| \leq (D-1) \left(2T + \dfrac{\tau_i}{w_i} -\dfrac{\tau_j}{w_j} \right).
\end{equation}
Here, $w$ is the priority weight of the function, and $\tau$ is it's average execution time in the interval.
For simplicity of design and analysis, we assume all functions have the same weight ($w=1$). 

The proof of this property hinges on the assumption that $queue.\VT < \GlobVT + \T$, which we meet in line~\ref{lst:line:filter}.
The original MQFQ picks a random \emph{arbitrary} queue meeting this criteria, whereas our MQFQ-Sticky adds an additional sorting criteria. 
Thus the possible space of dispatch decisions of MQFQ-Sticky is a subset of MQFQ, and we retain its theoretical fairness properties.
Because we select based  on longest queues (and lowest in-flight), a tighter bound may be possible for MQFQ-Sticky, but that is beyond the scope of this work. 


\algnewcommand{\LineComment}[1]{\State \(\triangleright\) #1}
\begin{figure}[t]
\begin{algorithm}[H]
\caption{\QName~algorithm.}
\label{algo:dispatch}
\begin{algorithmic}[1]
\Procedure{Dispatch}{}
\State $\GlobVT \gets min_{f \in \text{queues}}(f.\VT)$
\State $chosen \gets None$
\For{$queue \in queues $}
\State $update\_state(queue, \GlobVT)$ \label{lst:line:update_vt} 
\EndFor
\State $cand \gets filter(queue.active \wedge queue.len >0 \wedge queue.\VT < \GlobVT + \T, queues)$ \label{lst:line:filter}
\State $sort\_descending(cand, on: length)$
\If{$\D \ne 1$}
\State $sort(cand, on: in\_flight)$ \label{lst:line:in_flight}
\EndIf
\State $chosen \gets top(cand)$
\State $token \gets get\_\D\_token(chosen)$
\If{$token == None$}
\State \Return $None$
\EndIf
\State \Return $chosen.pop()$ \label{lst:line:done}
\EndProcedure
\\
\LineComment{Update state of queue, given the global \VT}
\Procedure{update\_state}{queue, \GlobVT} \label{lst:line:update_state}
\If{$queue.is\_empty~\textbf{and}~queue.in\_flight==0$}
\If{$Date.Now() - queue.last\_exec \ge TTL$}
\LineComment{Queue has expired}
\State $queue.state \gets Inactive$
\EndIf
\ElsIf{$queue.\VT - \GlobVT < \T$}
  \LineComment{Queue has exceeded threshold}
  \State $queue.state \gets Throttled$
\Else
\State $queue.state \gets Active$
\EndIf
\EndProcedure
\end{algorithmic}
\end{algorithm}
\end{figure}

\subsection{Integrated Memory Management and Scheduling}
\label{sec:design-cont-shim}

We use MQFQ queue states to guide memory movement.
When some queue becomes active, all its CUDA-malloc'ed regions are \emph{prefetched} into the GPU memory, in anticipation of use. 
We introduce and maintain a \emph{container pool} of such created GPU containers, and executions of the function results in a \emph{GPU-warm} start.
Each container has a custom shim that intercepts calls to the GPU driver, specifically those for initialization and memory allocations.
Requests by functions to allocate physical memory are captured in and converted into UVM (virtual device memory) allocations, allocation metadata is stored, and the result is returned to function.

The efficacy of this container pool is restricted by physical GPU memory, and thus throttled and inactive queues have their regions marked for eviction.
This entails swapping and moving their GPU memory regions back to the much larger host memory, with this eviction done asynchronously using LRU (least recently used) order.
In rare cases, a throttled and swapped out queue may get invoked again, which leads to a \emph{\quotes{GPU-cold but host-warm}} start since the container is already fully initialized, but data dependencies are not located on device. 
In the above case, prefetching may need to evict some other container's GPU regions, increasing latency. 


\subsection{GPU Load Management and Control}
\label{sec:gpu-man}


Dynamically setting \D~is dependent on the computing and memory resource limitations of GPUs.
The GPU monitor is responsible for two key mechanisms: tracking GPU assignment and monitoring GPU utilization. 
Creating a GPU context uses physical memory we can't control, so the monitor only allows a fixed number of containers to exist at one time.

Because we intercept memory allocations, we can closely track device memory usage of containers thanks to our driver shim, and only allow a new dispatch when the needed container won't overload the device's physical memory. 
Ensuring there is available compute is not as straightforward to manage as memory due to unpredictable function characteristics.
An ML inference task for example could have known compute usage, since input and weight tensors have fixed uses based on the execution graph.
Other types of GPU functions can launch compute kernels unpredictably, based on the application's internal control queue.
The actual size and number of kernel launches often vary with function arguments, making a priori utilization intractable.
Accepting this, we choose to externally monitor device utilization and launch new invocations when there is sufficient headroom for a new dispatch, assuming that it will consume $1/D$ GPU resources. 

The max concurrency $D$ is a tunable parameter.
As we shall show in the evaluation, increasing $D$ naturally increases utilization, but also increases contention and execution time. 
Since functions are small an unlikely to use entire GPUs, one possible design is one in which users specify the max GPU concurrency they are willing to accept, to control their quality of service. 
GPU servers in our FaaS pool belong to different $D$ classes, and functions are dispatched by the load balancer using consistent hashing~\cite{faaslb-hpdc22} to the appropriate server based on locality and resource availability. 
Other load balancing schemes may be possible using apriori function profiling to ``pack'' functions of different sizes.
Our focus with MQFQ-Sticky is to develop a general single-server scheduling solution decoupled from the cluster-level load balancing.

\vspace*{-0.2cm}


\section{Implementation and Microbenchmarks}
\label{sec:impl}


The architecture of MQFQ-Sticky requires the FaaS control plane to implement per-worker queues.
Because it already supports rudimentary GPU containers with NVIDIA-Docker support, and because of its per-worker queue architecture, we implement MQFQ-Sticky as a module inside Iluvatar~\cite{fuerst2023iluvatar}.
Iluvatar's lower overheads compared to OpenWhisk also allow us to evaluate GPU-function latency without additional system noise. 
Our GPU scheduler module is implemented in around 3,000 lines of Rust. 


Invocations are dispatched by a dedicated thread which monitors available GPU resources, and selects a function to execute based on Algorithm~\ref{algo:dispatch} described in the previous section. 
The dispatch thread is also notified upon invocation completion events, which helps maintain high device parallelism (\D). 
For servers with multiple GPUs, we maintain a single dispatcher which allows late binding of functions to individual GPUs.
Locality is still maintained by the dispatcher implementing \quotes{sticky} load balancing among GPUs: line~\ref{lst:line:in_flight} of Algorithm~\ref{algo:dispatch} helps us avoid moving functions across GPUs and reduce cold-starts.
Function queues and \VT~tracking are kept in one data structure, and protected behind Read/Write locks. 

\mhead{Utilization monitoring}
We track both GPU compute utilization and per-container and total GPU physical memory usage.
Using NVML~\cite{nvml}, we query utilization (every 200ms) and record its instantaneous and moving average, used for enforcing \D.
To track memory usage, our shim includes a report of memory allocations still held by the function to the worker alongside other invocation results.
This allows us to track the average memory used by a function, and determine memory pressure. 
Containers from inactive queues are evicted based on LRU order. 

\begin{figure}
  \includegraphics[width=0.38\textwidth]{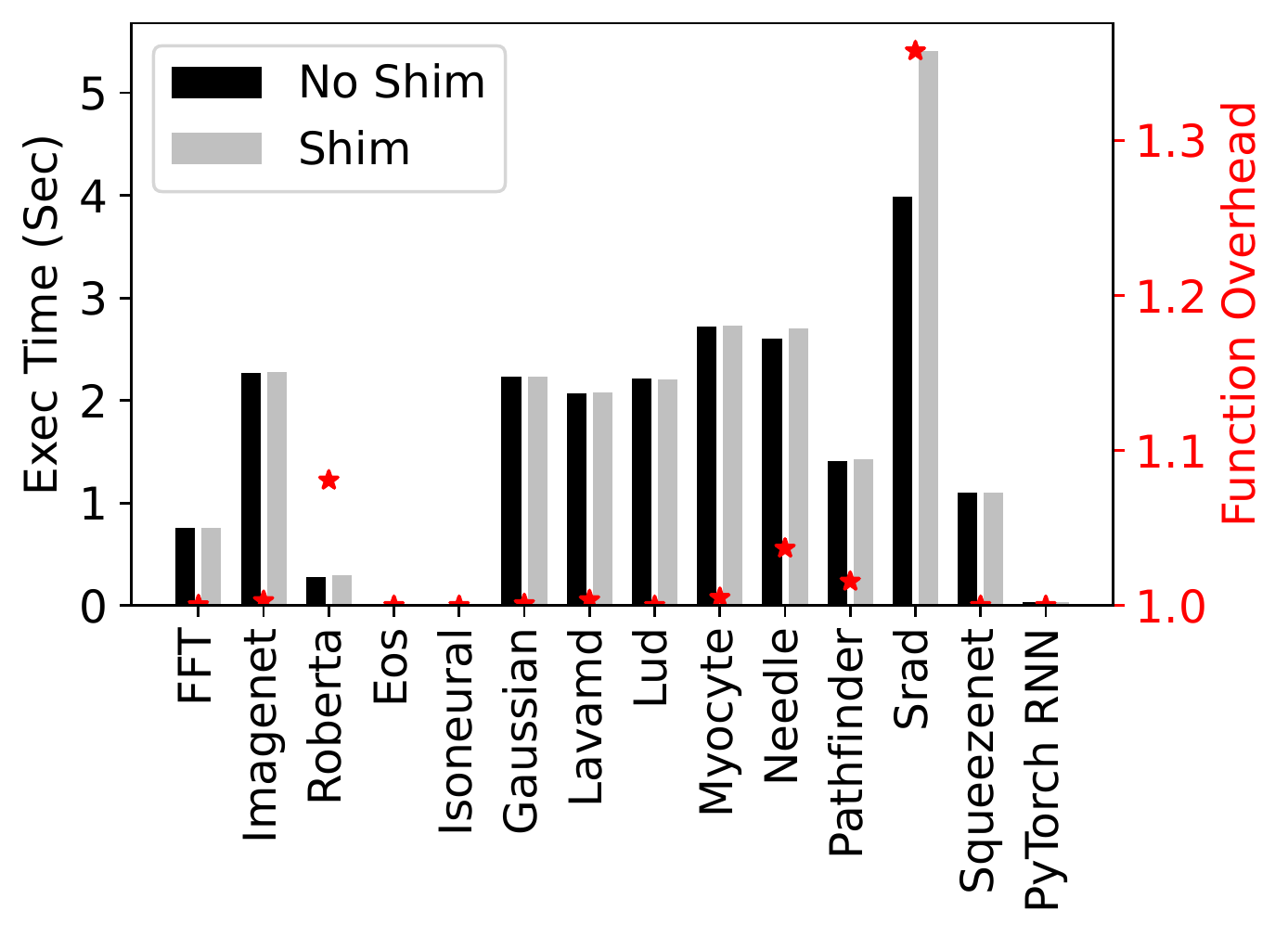}
  \vspace*{-0.15cm}
  \caption{Functions see little to no impact from our interception and substitution of allocation calls---in line with performance of general UVM applications.}
  \label{fig:shim-overhead}
  \vspace*{-0.3cm}
\end{figure}

\subsection{CUDA Interposition Shim}
We run functions inside Docker containers~\cite{docker-main} using the NVIDIA Container Toolkit~\cite{nvidia-container} to attach specific GPUs to them. 
For dedicated GPUs with limited memory, we use a CUDA interposition ``shim'' which intercepts CUDA calls made by the function. 
Our shim implementation is similar to NVShare~\cite{alexopoulos2023nvshare}, but simpler: we only use it for intercepting memory allocation calls and forcing the function to use virtual memory.
This requires about 500 lines of code (written in C) to be injected using \funcname{LD\_PRELOAD}.
This addition of a small additional ``driver interception'' module is analogous to the FaaS HTTP agent inserted by all FaaS providers for function invocation and tracking, and preserves the black-box requirement, since it requires no change to user-provided function code.

We use CUDA's Unified Virtual Memory (UVM) to oversubscribe device memory.
UVM uses a unified host-device memory address space, with memory pointers for applications being valid in both contexts.
The CUDA driver moves and ensures coherency of UVM memory between the host and device as use and pressure demands, mimicking disk-based swap space found in operating systems.
Our shim intercepts all calls to the driver for allocations for physical memory made via \funcname{cuMemAlloc}, and makes a UVM allocation of the same size using \funcname{cuMemAllocManaged}.
We record the size and memory pointer position, then return the pointer to the function, thus maintaining  execution transparency. 
It can use this memory as if it were physically allocated, reading, writing, or copying it to the host using traditional driver calls. 
If the function already uses UVM, we intercept and forward allocations, recording the returned metadata for our memory management tasks. 

The performance overhead of our interception is primarily influenced by the memory access patterns of the function and the extra layer of virtual memory (UVM), and is shown for different functions in Figure~\ref{fig:shim-overhead}. 
All results are averaged over 10 trials, and we see a negligible latency impact on most functions. 
The rest see single-digit percentage increases, with \funcname{Srad} standing out with a 30\% overhead in execution time due to the UVM shim. 
These results are in line with NVIDIA's own reporting on the performance change when migrating applications to UVM~\cite{nvidia-uvm}.
This low overhead shim is thus a suitable mechanism for overcommitting GPU memory. 

\begin{figure}
  \includegraphics[width=0.3\textwidth]{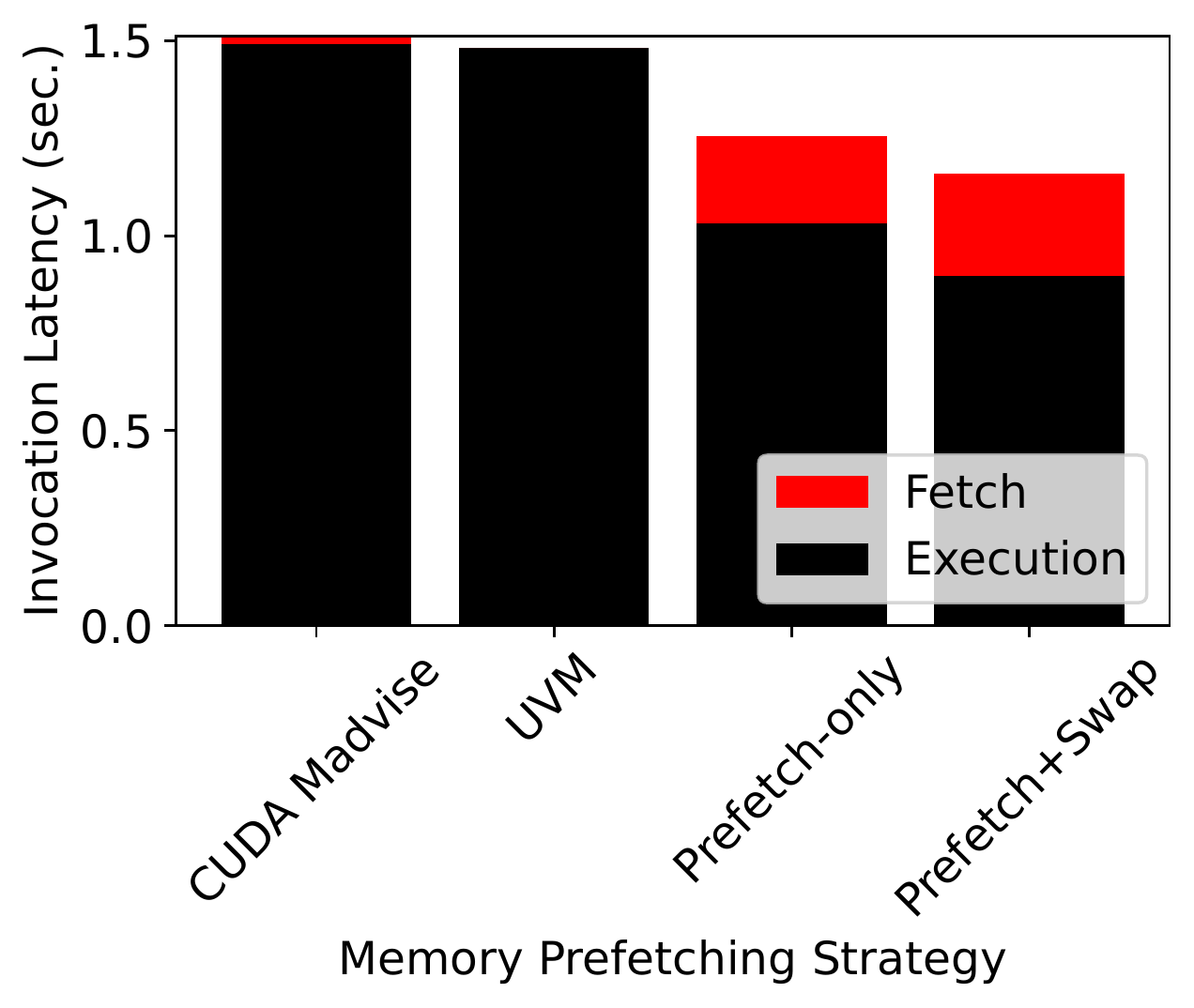}
  \vspace*{-0.15cm}
  \caption{Active memory management (\texttt{Prefetch+Swap}) improves execution latency.}
  \label{fig:mem-prefetch}
  \vspace*{-0.3cm}
\end{figure}

\subsection{Memory Management}

We implement the warm pool, memory prefetching, and swapping optimizations inside the control plane, integrated with the scheduler. 
We prefetch GPU memory of active functions, which is not provided out-of-the-box by CUDA UVM. 
Default UVM only moves memory on-demand, and also exposes \texttt{madvise} hints (via \texttt{cuMemAdvise}) for memory ranges, but neither allow for deterministic control of memory placement.
Our policy (\texttt{Prefetch+Swap}) asynchronously copies memory to the device before invocation, and swaps it back to host memory after the queue becomes idle (or evicted on-demand using a least recently used policy). 
When a queue becomes active or after it's chosen for dispatch, we direct our shim to prefetch a container's GPU memory using \funcname{cuMemPrefetchAsync}.
Doing this in a non-blocking manner allows us to overlap prefetching with the control plane marshaling invocation arguments to send to the container. 
Not having to block while waiting for memory to be moved saves significant time on the critical path.
When a queue is throttled or memory is needed to run other functions, we direct the shim to again use \funcname{cuMemPrefetchAsync} move memory off the device and into host memory.

We compare different memory management policies in Figure~\ref{fig:mem-prefetch}.
We run 16 copies of the FFT function from Table~\ref{tab:gpu-cpu}, each using 1.5 GB of device memory which oversubscribes the GPU memory by 50\%. 
Each copy is sequentially invoked 20 times.
The impact of these different memory policies are displayed in Figure~\ref{fig:mem-prefetch}, with average time spent in-shim shown in red and function code execution in black.
With such high overcommittment and the stock UVM driver handling data placement, the  execution time is 40\% worse than the optimal seen in Table~\ref{tab:gpu-cpu}.
Execution time is higher because memory must be paged in on-demand from the host as kernels access it, and old memory paged out.
Surprisingly, using \texttt{CUDA Madvise} to control memory placement performs slightly worse. 
Madvise doesn't move any memory and wastes time sending memory directives, with no benefit to execution time. 
We also implemented a \texttt{Prefetch-only} policy which does not remove function memory, relying on UVM to reclaiming pages. 
In contrast, our \texttt{Prefetch+Swap} policy reduces latency by over 33\% compared to stock UVM.
This shows that adding the async swapping optimization (our default) matches the ideal non-UVM execution time listed in Table~\ref{tab:gpu-cpu}. 



\section{Experimental Evaluation}
\label{sec:eval}

We use the \QName~implementation from the previous section and experimentally evaluate the resulting function performance in this section. For brevity, we shall also refer to it as ``MQFQ''.

\begin{figure*}
  \centering
\subfloat[GPU service time as functions are added to the workload at the 5 minute mark. MQFQ is fair, and provides all functions similar service, unlike popular functions dominating with FCFS. \label{fig:fair-micrhbench}]
{\includegraphics[width=0.3\textwidth]{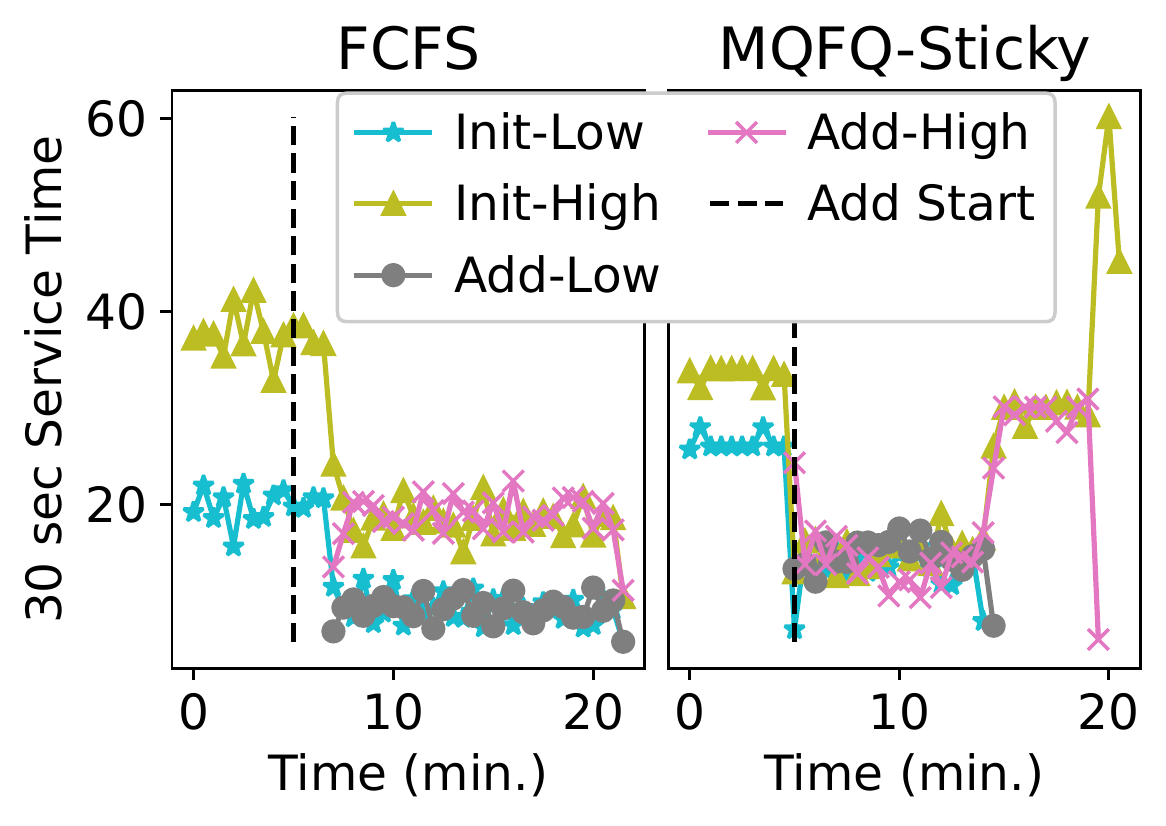}}
\hfill
\subfloat[The maximum difference in GPU execution time among all functions is significantly smaller than the theoretical upper-bound.  \label{fig:zipf-fair}]
{\includegraphics[width=0.3\textwidth]{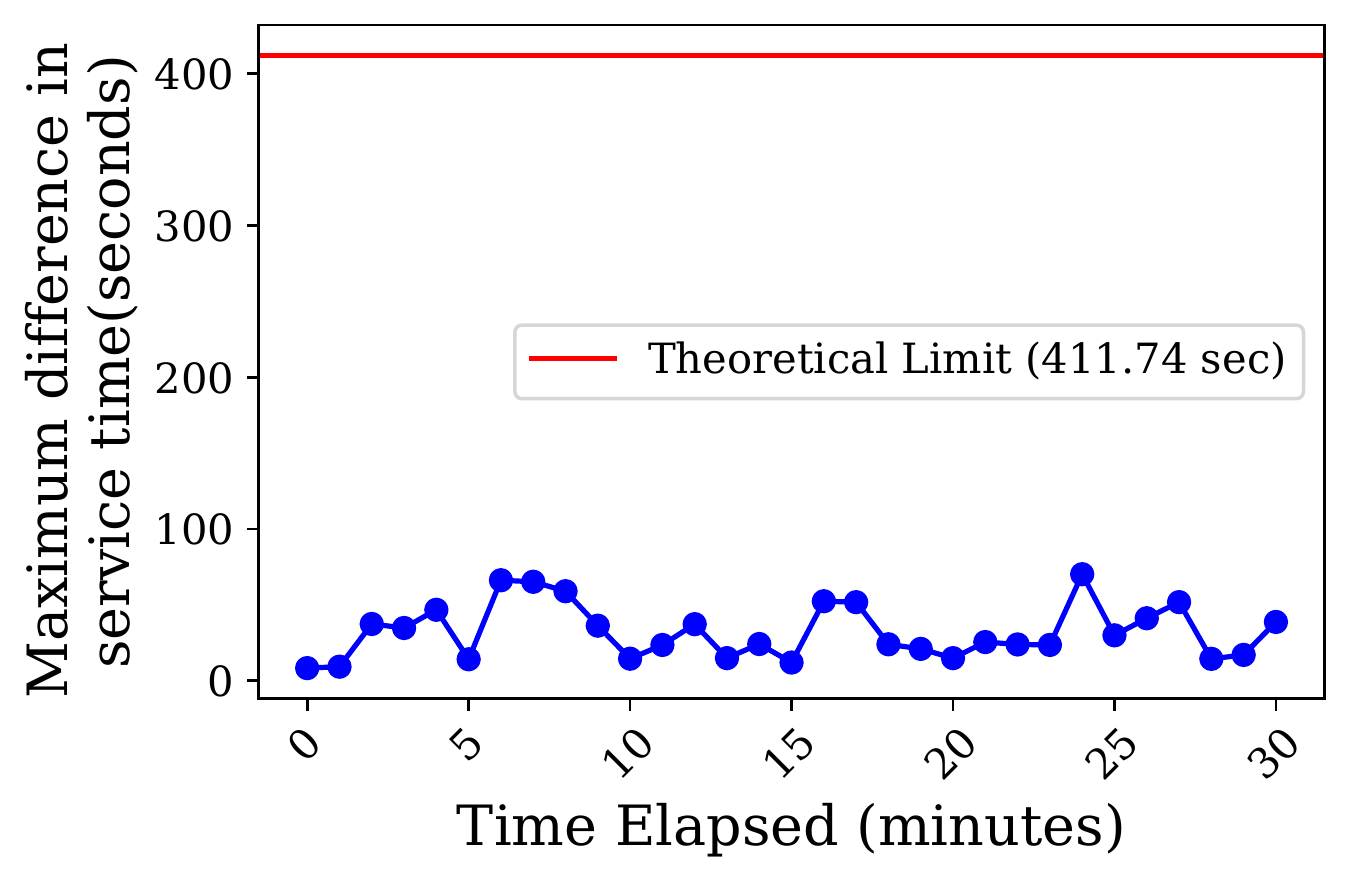}}
\hfill
\subfloat[MQFQ provides lower end-to-end latency across a range of workloads. \label{fig:zipf-load}]
{\includegraphics[width=0.3\textwidth]{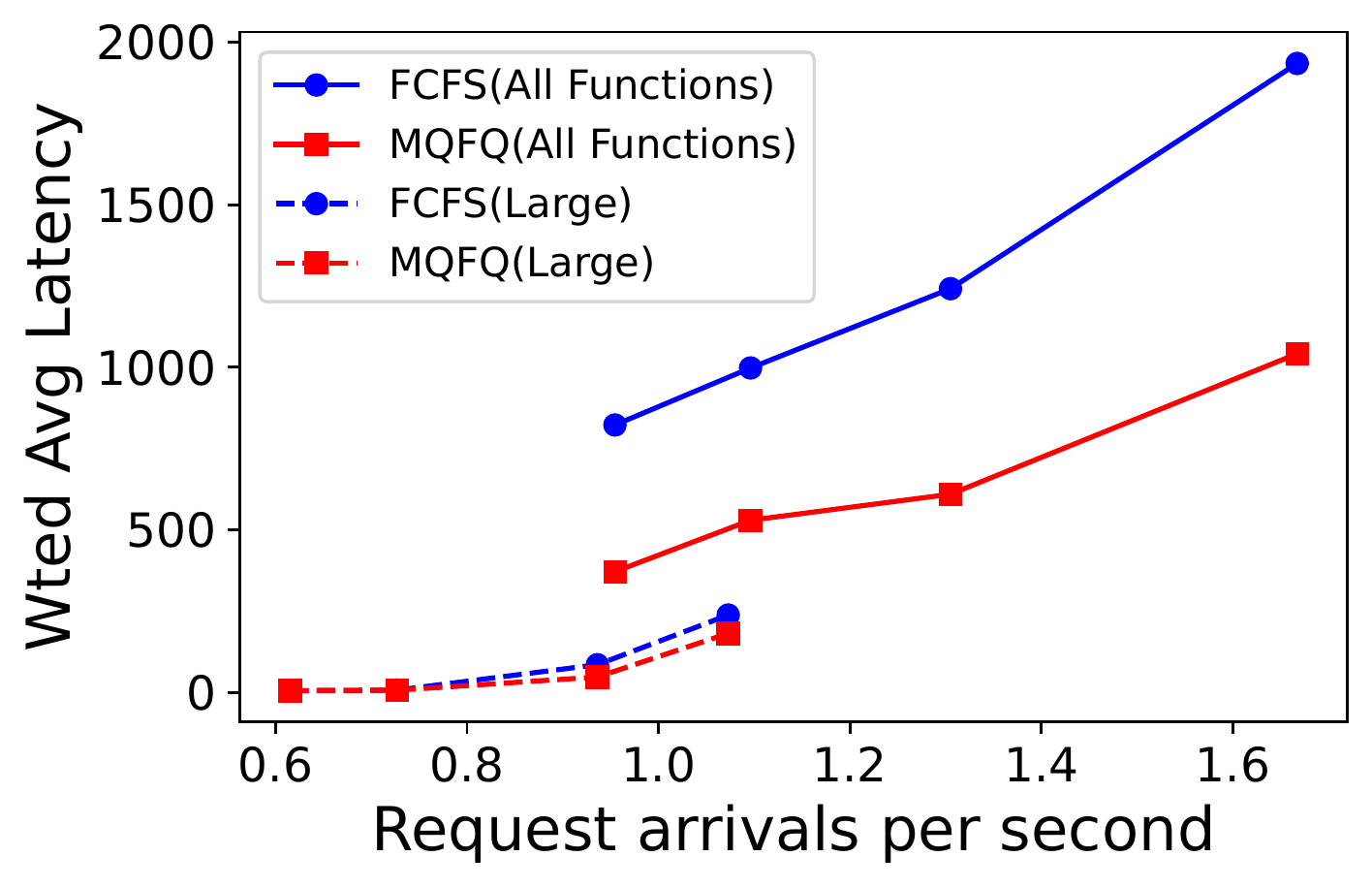}}
\label{fig:fair-all}
\caption{MQFQ-Sticky provides fair GPU access to functions, and also significantly reduces end-to-end latency.}
\vspace*{-0.4cm}
\end{figure*}



\mhead{Queueing Policies}
To examine the locality, fairness, and utilization tradeoff, we implement and evaluate several \emph{additional} queueing policies. 
The simplest is \fcfs, in which invocations executed in order of arrival.
In the \batch~policy, we insert invocations into per-function queues, and dispatch the entire \emph{queue} containing the oldest item---analogous to continuous batching used in modern LLM serving~\cite{sheng2024fairness}. 
Unlike \QName, \batch~greedily maximizes batch sizes (and hence locality) irrespective of the build-up of other functions.  
We compare against the scheduling approach from one the state of the art GPU multiplexing systems, Paella~\cite{ng2023paella}, which uses a Shortest Job First (SJF) policy with a fairness limiter. 
Paella schedules individual kernels from within GPU jobs, choosing the kernel with the expected shortest running time.
We adapt and reimplement its scheduling approach, and choose the shortest \emph{function}, running the invocation to completion.
All these policies use our memory management optimizations (warm pool, prefetching, etc.), and CUDA interposition shim, which allows for a pure queueing-policy comparison. 

\mhead{Setup and Workloads}
We use two different hardware platforms to show the effectiveness and adaptability of our design.
The first is an older Nvidia V100 GPU on a local testbed server with a 48 physical core Intel Xeon Platinum 8160 CPU and 250 GB of RAM. 
The second is a newer NVIDIA 24GB A30 on a Cloudlab server with two 16-core AMD 7302 CPUs and 128 GB of RAM.
The servers run Ubuntu 20.04 on kernel version 5.4 and Nvidia GPU driver version 470.239.06.
We restrict our evaluation to a single server, since load balancing does not play an important role for low arrival rate of GPU-function workloads. 

On the V100, MIG is unavailable, and MPS support is brittle and crashes the driver (and the server)---requiring us to evaluate performance without hardware multiplexing features. 
The use of older and non state-of-the-art GPUs will be increasingly important for cloud providers for sustainability, since it reduces the effective embodied emissions by increasing the hardware lifespan~\cite{acun_carbon_2022, jiang2024ecolife}.

MQFQ-Sticky has several parameters such as \T, \D, etc.
We conduct a detailed sensitivity study of these parameters in Section~\ref{sec:queue-knobs}, which shows that performance is ``stable'' across a large range of these parameters.
We therefore use the default parameters of $D=2, T=10, \alpha=2$, and warm-pool-size=32. 


All of our workloads are heterogeneous, and use functions from Table~\ref{tab:gpu-cpu}.
We create \emph{multiple} copies of the same function code, each with their own arrival rates. 
Each experiment is run with open-loop traces composed of 24 functions, and our results show the average of 5 repeated runs. 
We use two workload classes: Zipfian and Azure.
For the Zipfian workload, the inter-arrival-times of each function are exponentially distributed, and the average arrival rates of different functions are zipfian (parameter=1.5). 
This represents the widely used class of web and ML-inference workloads (e.g.,  ~\cite{gujarati2020serving}). 
For the Azure workload, we sample and scale the IAT distribution from the widely used Azure trace~\cite{shahrad2020serverless}, as is commonly done for FaaS evaluation~\cite{ustiugov2021benchmarking, ustiugov2023enabling}.
The original Azure trace is extremely heavy-tailed and is dominated by extremely short-running CPU functions.
We therefore use the above approach and use multiple samples, which yields different function mixes and invocation frequency distributions, shown in Table~\ref{tab:scaling}.

\begin{table}
  \addtolength{\tabcolsep}{-3pt}
  \begin{tabular}{@{}c|ccccccccc@{}}
    \hline
    Trace ID & 0 & 1 & 2 & 3 & 4 & 5 & 6 & 7 & 8 \\
    \hline
    Req/sec & 1.12 & 1.69 & 1.94 & 4.26 & 2.69 & 2.57 & 2.55 & 1.79 & 1.12 \\
    GPU Util (\textbackslash \%) & 37.9 & 44.3 & 48.8 & 67.0 & 77.1 & 43.2 & 79.9 & 44.9 & 54.2 \\
  \end{tabular}
  \caption{We use multiple samples of the Azure trace~\cite{shahrad2020serverless}, to evaluate on workloads with different function and invocation characteristics.}
  \label{tab:scaling}
  \vspace*{-0.5cm}
\end{table}

\subsection{Fairness and Latency}

\mhead{Service-time Fairness}
This subsection uses the Zipfian workload class.
We start by showing that the device service-times provided by \QName~are fairly/equally distributed across functions.
In this experiment (Figure~\ref{fig:fair-micrhbench}), we run four copies of a single function (\texttt{cupy}), two \quotes{Low} functions have an IAT of $\alpha$ and two \quotes{High} functions have and IAT of $2*\alpha$.
For each function, its service time (execution time on GPU) over 30-second intervals is shown.
Initially, both functions get high service, which drops after the other two functions are introduced. 
With FCFS, the popular functions dominate and get more service. 
In contrast, MQFQ provides the same service to all four functions---demonstrating the key fairness property (equal service to all backlogged functions). 

Next, we take closer look at MQFQ-Sticky's fairness behavior across time.
For the full Zipf workload with 24 functions, we examine MQFQ's ``unfairness'' and ability to meet the theoretical fairness bound of Equation~\ref{eq:fairness}. 
Figure~\ref{fig:zipf-fair} shows the maximum difference between the service received by functions in a 30 second period, which is essentially $S_{\text{max}} - S_{\text{min}}$. 
We see that the average gap is less than 50, comfortably below the theoretical upper-bound of 411 obtained using Equation~\ref{eq:fairness}. 

\mhead{End-to-end Latency vs. Load}
MQFQ also provides significant improvement in the end-to-end latency which includes both the queueing and the service time.
The weighted-average latencies is equal to $\dfrac{\sum N_i L_i}{\sum N_i}$, where $N_i$ is number of invocations of function $i$ and $L_i$ is its average latency.
In Figure~\ref{fig:zipf-load}, we show this metric for many different Zipfian workloads.
With all 24 functions included, MQFQ reduces the latency by more than $2\times$ across a range of input loads (total arrival rate).
The high latency is largely a result of the scheduling and cold-start overheads due to the large working set.
With a workload comprising of only large functions (warm execution time>5 seconds), the latency drops significantly, and MQFQ improves upon FCFS by 15\% at the higher loads. 

\noindent \textbf{Result:} \emph{MQFQ provides fair access to GPUs to functions. It also reduces end-to-end latency compared to FCFS by $2\times$ at high load.}

\vspace*{-0.3cm}
\subsection{Queueing-Policy Comparison}
\label{sec:queue-perf}

\begin{figure*}[h]
  \centering 
  \subfloat[\textbf{Average latency} is $2-5\times$ lower with \QName~for different device-parallelism (\D) levels.  \label{fig:queue-e2e}]
  {\includegraphics[width=0.32\textwidth]{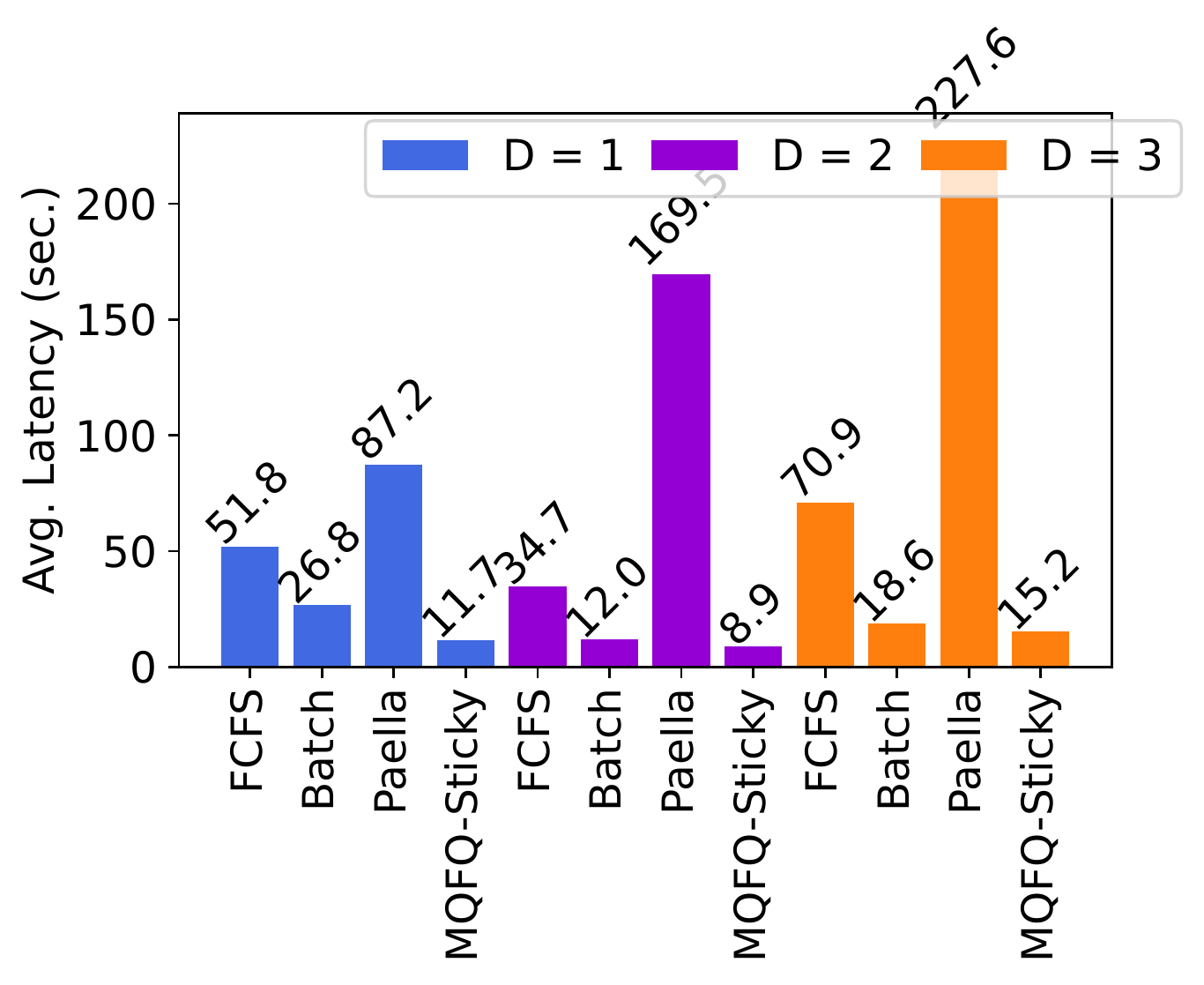}}
 \hfill  
\subfloat[The average and variance of \textbf{per-function latency} is much lower with \QName.  \label{fig:queue-fairness}]
{\includegraphics[width=0.35\textwidth]{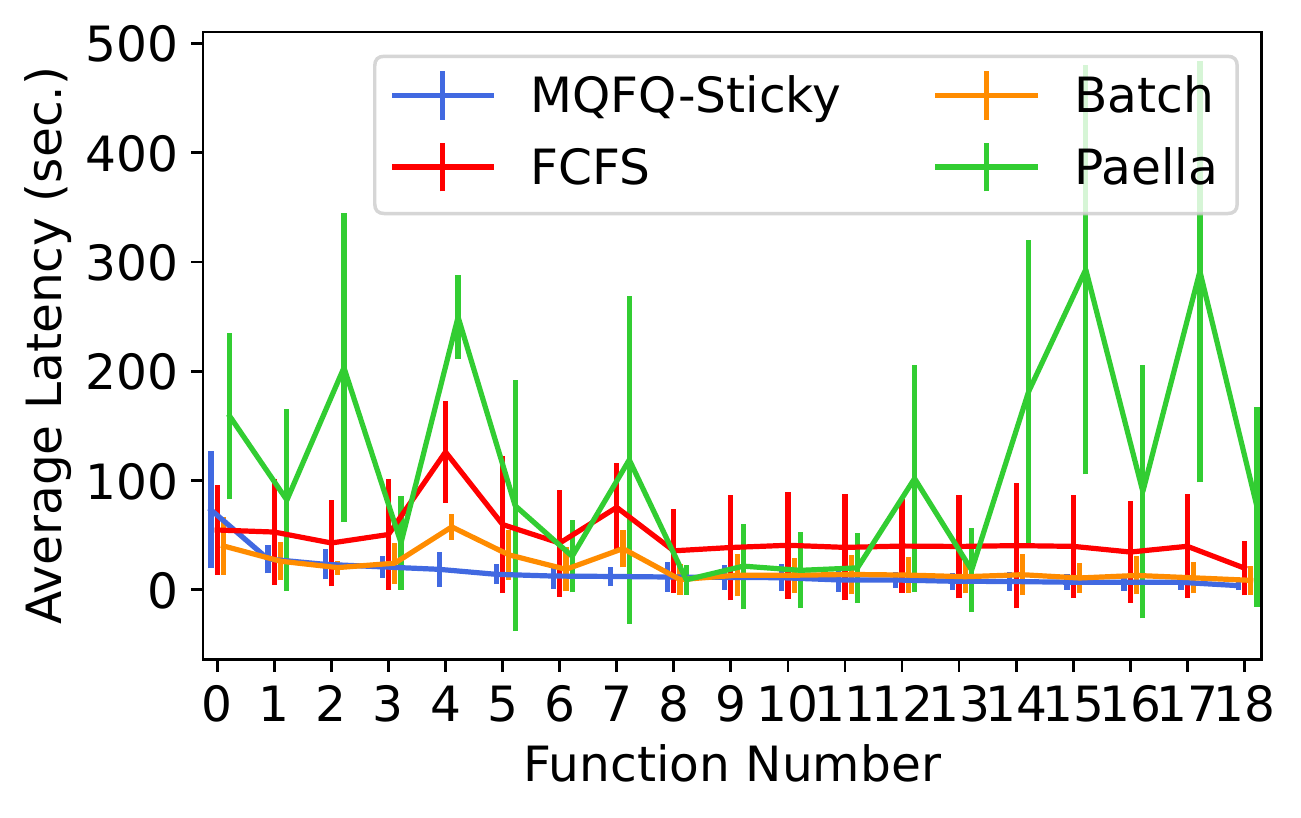}}
\hfill 
\subfloat[\textbf{Device utilization} for the medium-load trace. \label{fig:util} ]
{\includegraphics[width=0.3\textwidth]{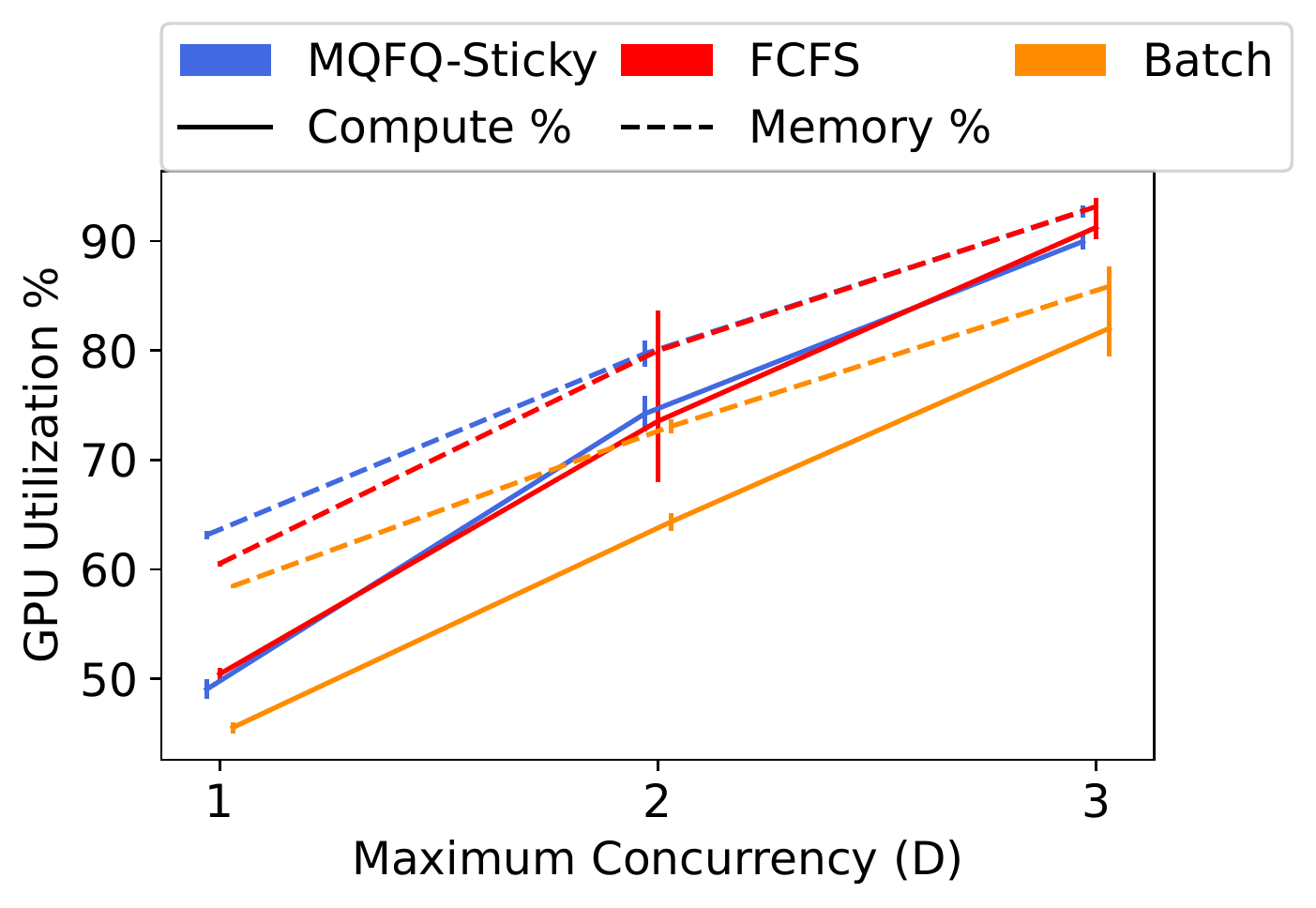}}
\label{fig:base-all}
\caption{Latency, fairness, and utilization for a medium-intensity FaaS workload.}
\vspace{-0.4cm}
%
%
\end{figure*}

To characterize the differences due to queueing policies, we show the empirical evaluation with a \emph{medium-intensity} Azure workload (trace 4 in Table~\ref{tab:scaling}), which comprises of 19 functions. 
This workload results in average GPU utilization of around 70\% (Figure~\ref{fig:util}), and represents the average case. 

\noindent \textbf{Average Latency.}
The latency across all invocations is shown in Figure~\ref{fig:queue-e2e}.
Not shown in the figure is the current baseline \naive~scheduling with nvidia-docker, which does not have a container pool and suffers from excessive cold-starts. 
\textbf{The \naive~average latency is close to 3,000 seconds---a $300\times$ overhead.}
The high latency is because of every invocation results in a cold-start, causing a large queue buildup.
Note that our workload trace is open-loop---with invocations generated at pre-determined timestamps.
\emph{Because the standard GPU-container overhead is so high, in this rest of this section, we will retain our memory-management optimizations when comparing MQFQ-Sticky with other queueing policies.}

A D=1, when only one function is serviced at a time, MQFQ approximates classic SFQ~\cite{goyal1997start}, and outperforms \fcfs~by $5\times$ with a 11.8 vs 51.8-second average respective latency thanks to its locality and fairness oriented design. 
Paella's SJF encourages locality at the expense of long functions that experience head-of-line blocking, resulting in a $8\times$ to $20\times$ higher latency. 
\batch~has middle of the road performance, lacking fairness and advanced locality policies.
At higher concurrency levels, \QName~improves latency by an additional 25\% to an average of 8.9-seconds per invocation. 

Most policies benefit from higher GPU concurrency (D) which decreases the queue wait times. 
However for Paella, it degrades performance, because its SJF dispatching results in  concurrent invocations of the same function, increasing the number of cold starts~\cite{ristov2022colder}.  
When \D~is set too high (\D=3), the device cannot handle the higher concurrency, and all policies suffer varying degrees of degradation due to resource contention and interference.

\noindent \textbf{Per-function latency.}
In Figure~\ref{fig:queue-fairness}, we show the per-function latency (averaged across all its invocations).
\fcfs~has the worst global inter-function latency variance (752), and the highest average latency.
\QName~reduces latency in the range of $2-10\times$, and has only one-third the inter-function latency variance of \fcfs.
Also, the invocation latency variance for each function (the error bars) is $3-4\times$ lower compared with \fcfs~and \batch. 

\noindent \textbf{Result:} \emph{\QName~reduces average latency by $5\times$  across all functions, and also reduces their jitter and tail latency by $3\times$--$4\times$.}

\subsection{MPS, MIG, and Multi-GPU}

\begin{figure*}
  \centering
  \subfloat[Function latency normalized to \QName~without any GPU spatial multiplexing.
  \label{fig:a30-multi-trace}]
  {\includegraphics[width=0.32\textwidth]{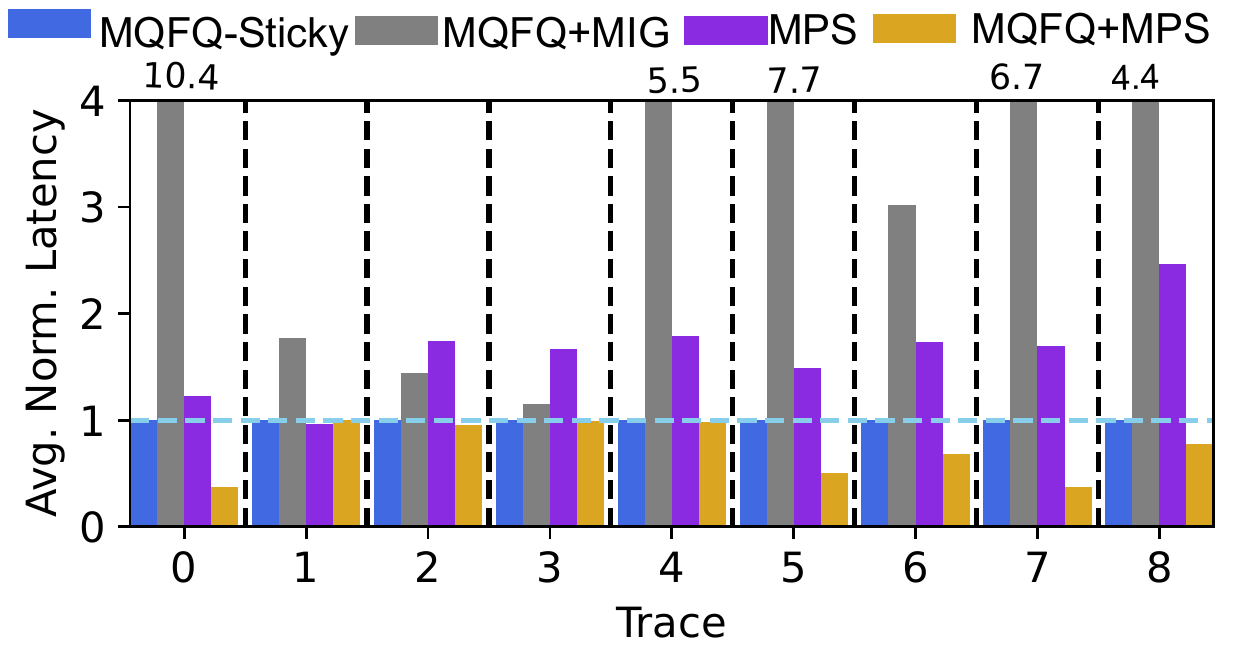}}
  \hfill
  \subfloat[MIG significantly degrades performance for a few functions such as RNN, SRAD, and FFT.
  \label{fig:a30-func-exec}]
  {\includegraphics[width=0.36\textwidth]{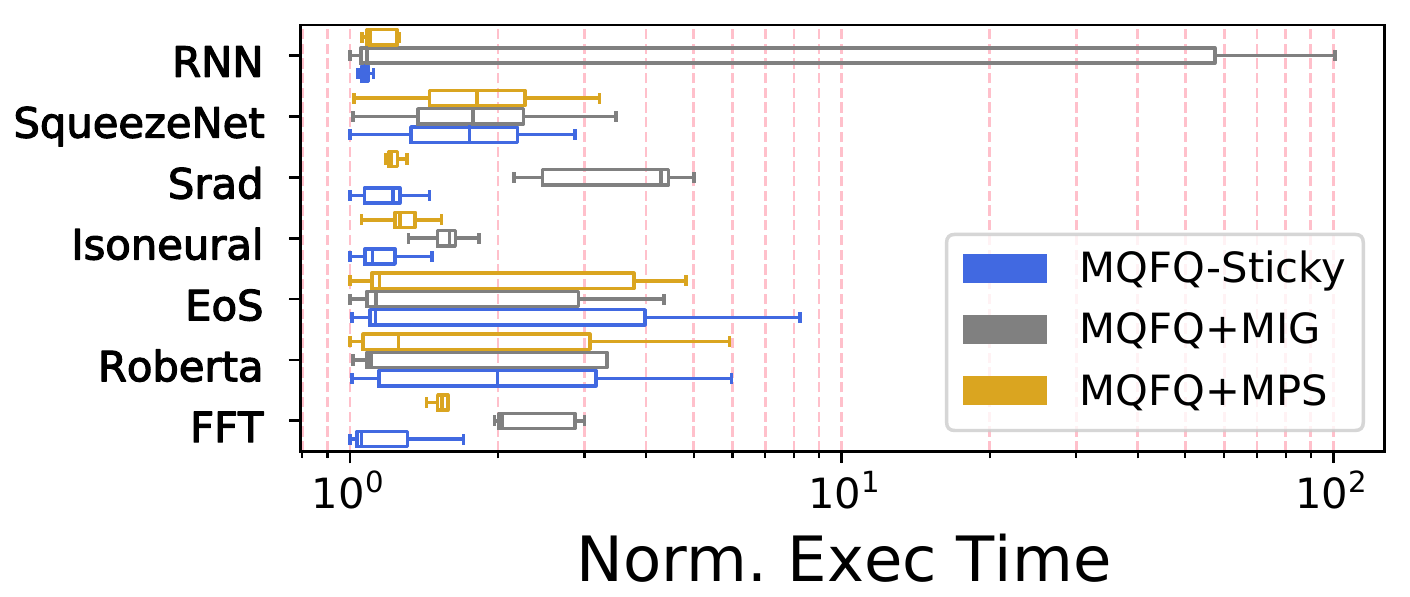} }
  \hfill
  \subfloat[\QName~also uses locality-aware scheduling for multiple GPUs, significantly reducing queuing.
  \label{fig:multi-gpu}]
  {\includegraphics[width=0.3\textwidth]{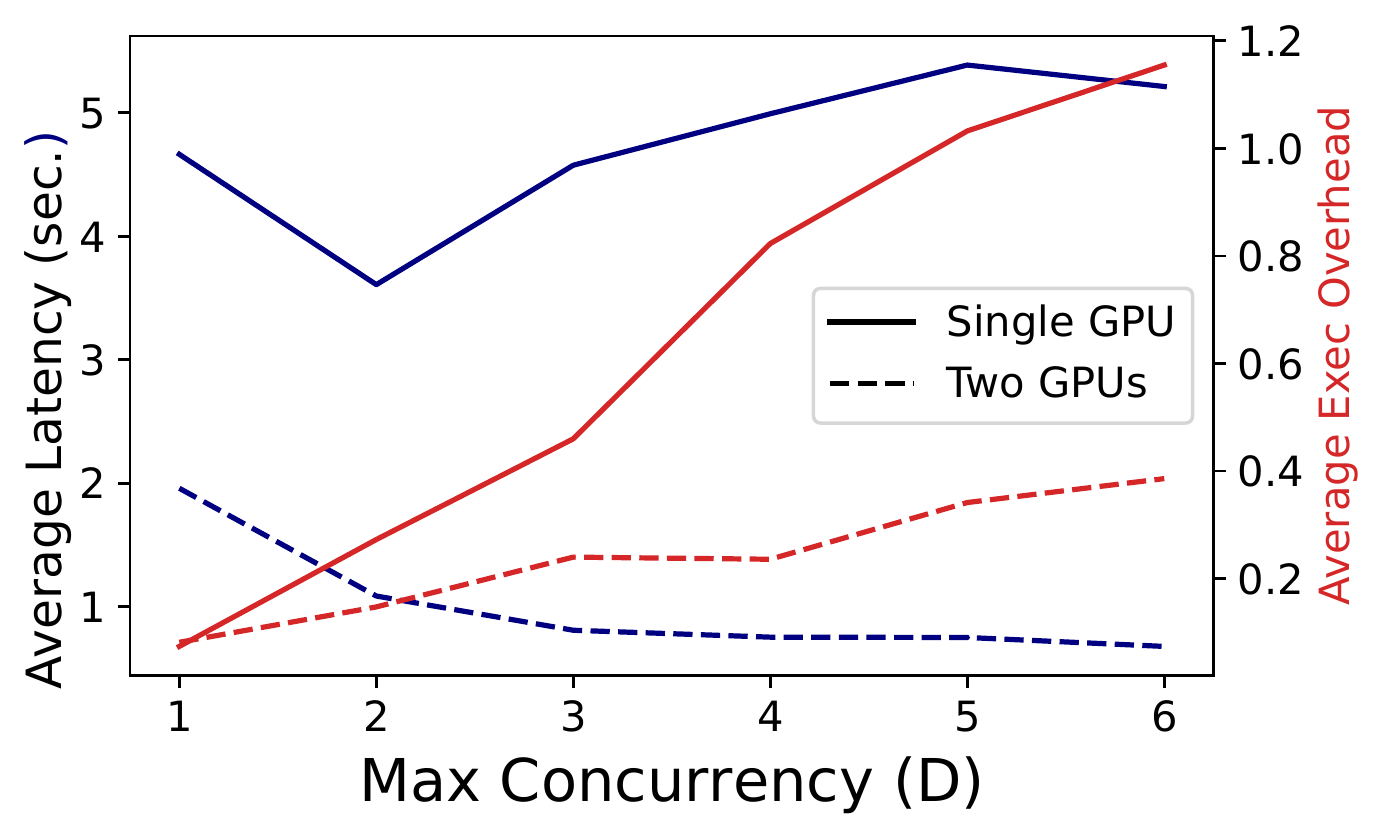}}
  \caption{Latency, fairness, and utilization for a medium-intensity FaaS workload.}

\end{figure*}

We now examine the effectiveness of MQFQ with MPS and MIG on the A30 GPU. 
Figure~\ref{fig:a30-multi-trace} compares the latency across different Azure-sampled workloads, with the weighted average latency normalized to MQFQ-Sticky without any of these features. 
We create two \textbf{MIG} slices, and MQFQ+MIG schedules functions across these two ``vGPUs''.
Surprisingly, this increases latency, and primarily occurs because of higher execution times for certain functions running on the smaller MIG slices, shown in Figure~\ref{fig:a30-func-exec}. 
Note that the same unmodified functions run on the MIG slices without accounting for the reduced resource availability.
This causes some functions such as RNN and FFT to see a large slowdown with MIG---increasing the average latency. 

Turning back to Figure~\ref{fig:a30-multi-trace}, we now look at the \textbf{MPS} performance. 
With pure MPS without MQFQ, the latency increases by 3 to 240\% compared to \QName.
However, when MPS is used \emph{with} MQFQ, we can get the best of both worlds: MPS can schedule the individual kernels and thread launches to improve low-level throughput, and MQFQ provides the higher-level scheduling.

\noindent \textbf{Result:} \emph{MQFQ+MPS reduces latency by up to $80\%$, showing the versatility of our design: it can provide low latencies both with and without hardware multiplexing support.}

\mhead{Multiple GPUs}
Our system easily scales to orchestrating and dispatching across multiple physical GPUs. 
We run a high-load trace and show the comparison in Figure~\ref{fig:multi-gpu} after we add a second, identical, V100 GPU to the server. 
Two GPUs not only allows us to run $\D\times2$ invocations, but also do on-the-fly load balancing between them to avoid compute contention with higher \D. 
As a baseline, the multi-GPU blue dashed line has $2.3\times$ lower latency at \D=1.
At higher device parallelism, the multi-GPU case sees a latency reduction of $4\times$ vs. the single GPU setting. 
Device parallelism also slightly increases the execution overhead due to interference, but is offset by the smaller queues. 


\vspace*{-0.5cm}
\subsection{Impact of Scheduling Parameters}
\label{sec:queue-knobs}

In this subsection, we explore the effects of configuration knobs to see their effect on performance, which also sheds a light on the empirical relationships between fundamental parameters of locality and throughput.

\begin{figure*}
  \centering
  \subfloat[Larger T yields more batching and using historical function execution latencies helps significantly lower latency over classical queuing. \label{fig:T-service}]
  {  \includegraphics[width=0.3\textwidth]{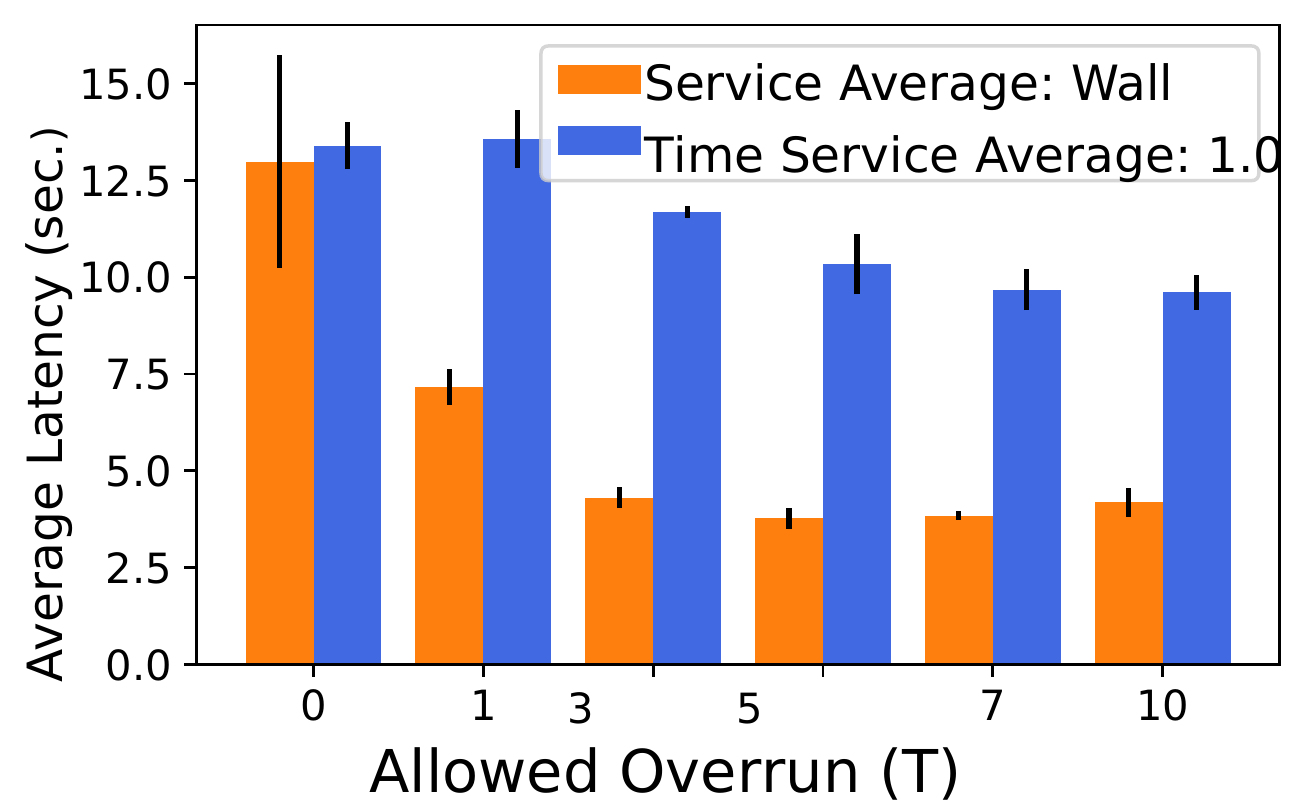}}
  \hfill 
    \subfloat[Anticipatory queue keep-alive (non-zero queue TTL) can reduce latency by up to 50\%. We use function IAT for scaling the TTL. \label{fig:queue-ttl}]
  {  \includegraphics[width=0.3\textwidth]{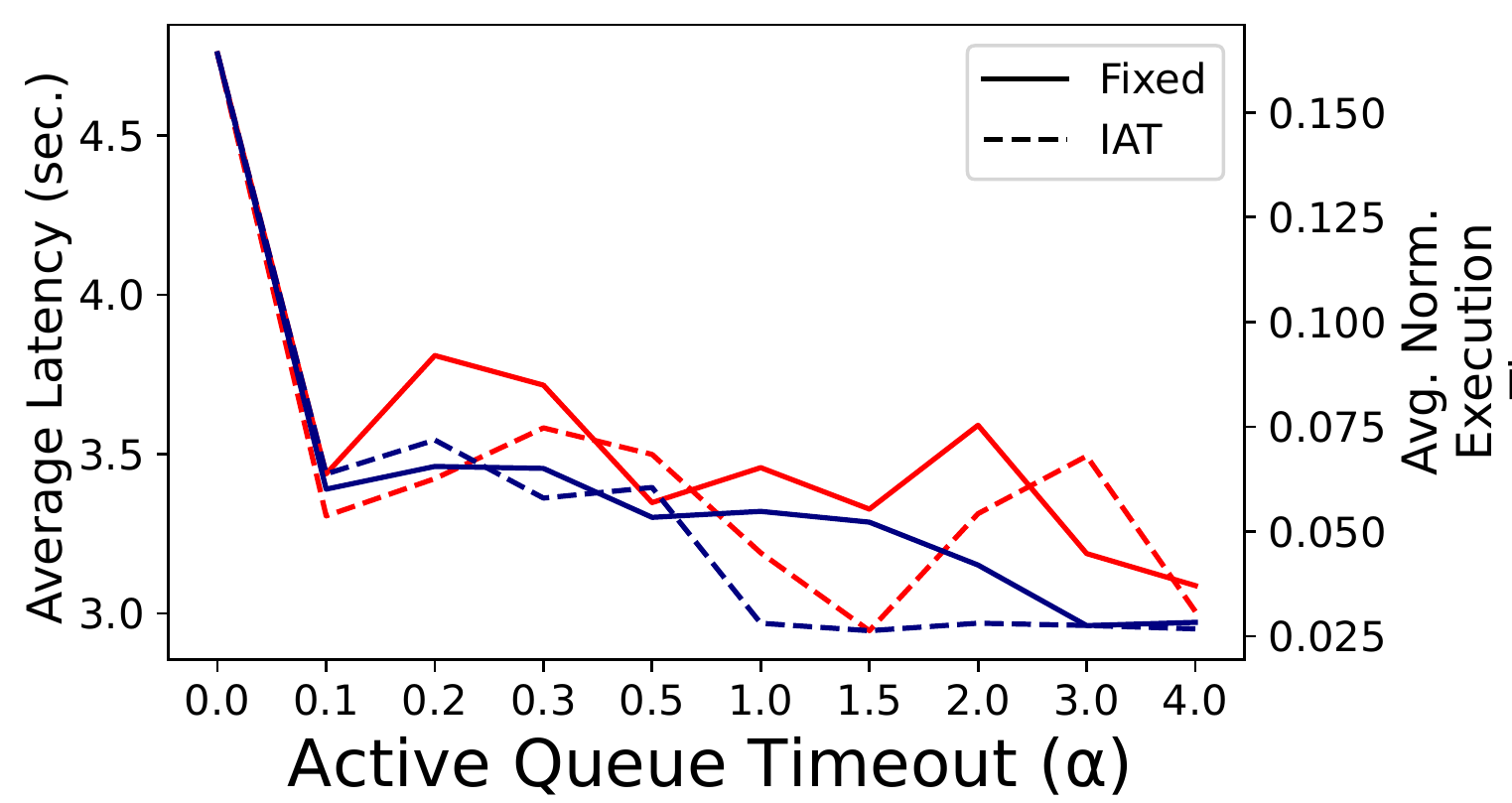}}
  \hfill 
  \subfloat[Container-pool reduces cold-starts. \QName~provides higher locality.  \label{fig:caching}]
  {  \includegraphics[width=0.3\textwidth]{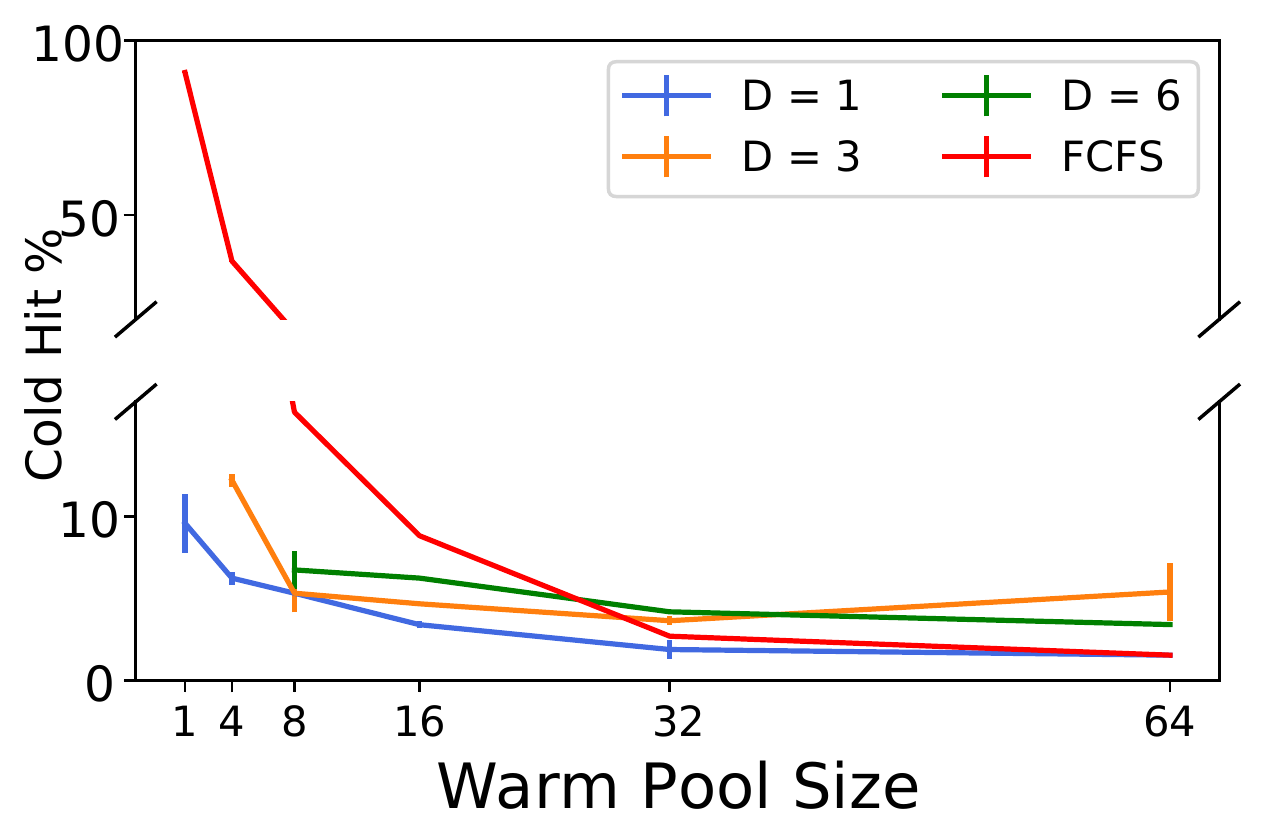}}
  \caption{Queue TTL and device parallelism (\D) are set based on workload and utilization.}
  \label{fig:knobs-all}
\end{figure*}

\mhead{Queue over-run (\T)}
The tradeoff here is that larger \T~results in more locality and batching opportunities, but decrease fairness, since queues may get to monopolize resources for longer before being throttled.
We examine this in Figure~\ref{fig:T-service}, which shows the average latency decreasing, but with diminishing returns, as the over-run is increased.
With strict fair queueing ($T=0$), the weighted average latency is $2.5\times$ worse, which shows the utility of the multi-queue approach.
Interestingly, the performance is not overly sensitive to $T>0$, and operators can choose from a wide range. 

\mhead{Function characteristics}
By default, we use per-function service times for adjusting the virtual timers (``wall time'' in Figure~\ref{fig:T-service}).
If we ignore function heterogeneity and assume all functions have the same run-time, (``1.0'' in the figure), then long-running functions can dominate.
The running-average execution times used by MQFQ-Sticky significantly improves the latency, by up to $2.7\times$, especially with  batching ($T>0$). 

\mhead{Queue keep-alive TTL}
We use anticipatory scheduling to keep empty queues ``active'' for a certain TTL, which depends on the function's inter-arrival-time ($\text{TTL}=\alpha \times \text{IAT}$). 
Figure~\ref{fig:queue-ttl} shows the improvement to both execution time as compared to ideal warm performance and latency as the TTL grows.
Without this anticipatory scheduling ($\alpha=0$), the average latency and execution time both increase by 50\%, which shows that it is a key mechanism for improving performance.
The execution overhead is reduced because of warm start locality, and is the main factor behind latency reduction. 
Setting any TTL (the solid line), at even a small 0.1 seconds, improves latency and overhead by 25\% and 50\% respectively. 
Increasing the TTL to up to 4 seconds sees significant, but diminishing returns.

Figure~\ref{fig:queue-ttl} also shows the effect of using per-function IATs, and compares against a fixed global IAT approach, where the TTL for all functions is the same.
Using the per-function IAT improves the latency by 15\% across a wide range of $\alpha=(1,3)$.
Higher TTLs ($\alpha > 3$) may result in too many active functions and pressure on the container pool. However, our design is robust to very large TTLs: the pool uses LRU eviction and the resulting impact on performance even at high TTLs is low.
\emph{Thus, anticipatory scheduling improves latency by 50\%, and \QName~performance is not sensitive to the TTL because of it's LRU managed container pool and proactive memory management.} 

\mhead{Container Pool Size}
For temporal locality, the invocation patterns and batching plays a key role in reducing cold starts.
The performance difference between \QName~and \fcfs~can be largely attributed to the cold-hit ratio of the invocations.
Figure~\ref{fig:caching} shows the \quotes{miss-rate curves} for the medium-intensity trace as we increase the number of containers in our container pool.
We show the number of containers in the pool, rather than MB of pool memory for simplicity.
Idle containers do take up CPU memory, and work managing the memory used by caching containers is orthogonal to our design~\cite{faascache-asplos21}.
Since \QName~prefers smaller batches of functions and does anticipatory keep-alive, it has a high temporal locality and its cold-hit \% is in the range of 2-8\% across a range of pool sizes and device concurrency.
In contrast, \fcfs~has 50\% cold-starts with a pool size of 4, and achieves parity with \QName~only at largest pool sizes when the popular functions can fit in the container cache.

The \textbf{preferential queue dispatch} in MQFQ-Sticky increases locality and prioritizes long queues. Without it, we observe a 1\%-30\% increase in latency. 
We also compared against the state-of-the-art CPU-specific earliest effective virtual deadline policy~\cite{fuerst2023iluvatar}, which also considers locality and load.  Compared to it, MQFQ-Sticky reduces latency by 40\% on average.
Recall that we apply our memory management and other optimizations to all the comparison policies, since without them, the latency overhead is more than $100\times$. 
Finally, we observed similar performance characteristics in a load-balanced cluster. Because the Iluvatar default is consistent hashing, it reduces the number of unique functions per server, and keeps the per-function traffic distribution the same.
Just as we achieve performance benefits from integrating server-level queueing and memory management, similar gains can be achieved in the future with integrated load balancing.

\noindent \textbf{Result:} \emph{Our key contributions such as a GPU container pool, queue over-run, anticipatory scheduling, and utilization-driven concurrency, all contribute to latency reduction by $1.5-10\times$ each. A wide range of these parameters yield similar performance, making our system robust, yet still providing operators enough flexibility for fine-tuning based on workloads and operational requirements.}


\vspace*{-0.4cm}
\section{Related Work}
\label{sec:related}

\noindent \textbf{Scheduling} is crucial for FaaS performance, with key tradeoffs in late vs. early binding~\cite{kaffes2021practical, kaffes_hermod_2022, serverless_sched_money_2024}.
For CPUs, queueing policies which prioritize function size~\cite{zuk_call_2022} can reduce latency, at the risk of stravation for longer functions.
In the offline setting with a limited number of batch jobs with known utilities,~\cite{mo_optimal_2024} also investigtes the efficiency and fairness tradeoffs in GPU scheduling.

\noindent \textbf{Locality} is an important design and optimization principle in FaaS---and is a fundamental result of code and data initialization required for each function.
Keep-alive policies for warm-starts apply temporal locality~\cite{roy2022icebreaker, ebrahimi2024cold, vahidinia2022mitigating, shahrad2020serverless} and caching~\cite{faascache-asplos21, sundarrajan2017footprint} principles for the CPU memory pool; load balancing also benefits from stickiness~\cite{package-cristina-19, faaslb-hpdc22, abdi2023palette}.
We extend these principles to GPU functions via locality enhanced fair queuing and proactive memory management.


\noindent \textbf{GPUs in serverless computing} is already a rich and fast-growing area of research. 
A big portion of prior work~\cite{naranjo2020accelerated, fingler2022dgsf, kim_gpu_2018} focuses on disaggregated accelerators, with GPUs accessed over the network using techniques such as rCUDA~\cite{duato2010rcuda}.
In contrast, we look at local GPUs without remote execution. 
Using FaaS-inspired abstractions to provide GPU acceleration as a service is also common: applications are broken down into kernels which can be run \quotes{anywhere}.
Kernel-as-a-Service~\cite{pemberton2022kernel} and Molecule~\cite{du2022serverless} are two examples of this approach, where the main challenges are designing and providing efficient and usable API-remoting mechanisms. 
\cite{juan_reducing_2023} also uses remote memory pooling to address the exacerbated cold-start problems for GPUs, and also proposes parallel data-dependency and compute context prefetching through code-level optimizations. 
Paella~\cite{ng2023paella} similarly breaks apart model inference tasks into CUDA kernel launches to minimize scheduling \quotes{bubbles}.
These and other recent~\cite{sage_zhao_towards_2024} specialized code-modifying techniques are orthogonal to our work, since we require general black-box functions.


The popularity of \textbf{ML inference} has resulted in a large number of specialized solutions to efficient GPU scheduling, which have similar challenges, but  different optimization spaces: inference resource requirements are much more deterministic~\cite{gujarati2020serving} and thus amenable to data-driven optimization~\cite{ali2022optimizing}, and the lack of isolation among requests provides many locality-enhancing and batching opportunities~\cite{yang2022infless, satzke2020efficient}. 
For instance, both FaST-GShare~\cite{gu2023fast} and TGS~\cite{tgs_wu2023transparent} leverage profiles of ML workloads to monitor GPU utilization and use 2D bin-packing (with time and memory dimensions) to schedule inference workloads.

\section{Conclusion}
\vspace*{\subsecspace}
We showed that harnessing GPU resources for functions is practical, and good performance can be obtained using integrated locality-aware scheduling and memory management.
Black-box containerized functions and heterogeneous and dynamic function workloads present many challenges to efficient GPU utilization. 
MQFQ-Sticky, our scheduling algorithm, is inspired by I/O fair scheduling. Empirical analysis of its performance indicates it reduces function latency by more than $20\times$ compared to other queueing policies and more than $300\times$ compared to unoptimized GPU containers.


\bibliographystyle{ACM-Reference-Format}
\bibliography{gpu-q-faas}

\end{document}